\documentclass[aps,prc,showpacs,twocolumn,superscriptaddress,nofootinbib,preprintnumbers,floatfix]{revtex4-1}%preprint
\usepackage{longtable}
\usepackage{subfigure}
\usepackage{mathrsfs}
\usepackage{graphics}
\usepackage{epstopdf}
\usepackage{hyperref}
\usepackage{amssymb}
\usepackage{amsmath}
\usepackage{array}
\usepackage{graphicx}% Include figure files
\usepackage{dcolumn}% Align table columns on decimal point
\usepackage{bm}% bold math
\usepackage{epstopdf}
\usepackage{tensor}
\usepackage{gensymb} 
\usepackage{cancel}
\usepackage{float}
\usepackage{color}
\usepackage{float}

\usepackage[normalem]{ulem}  % \sout{old text} for strikeout

\renewcommand\sout{\bgroup \color{red} \ULdepth=-.5ex \ULset}
\begin{document}


\title{${}_{\Lambda\Lambda}^{\,\,\,\,5}$H and ${}_{\Lambda\Lambda}^{\,\,\,\,5}$He hypernuclei reexamined in 
halo/cluster effective theory}


\author{Ghanashyam Meher}
\email[]{ghanashyam@iitg.ac.in}
\affiliation{Department of Physics, Indian Institute of Technology Guwahati, 781 039 Assam, India}
\author{Udit Raha}
\email[]{udit.raha@iitg.ac.in}
\affiliation{Department of Physics, Indian Institute of Technology Guwahati, 781 039 Assam, India}
%\thanks{}


%\date{\today}

%%%%%%%%%%%%%%%%%%%%%%%%%%%%%%%%%%%%%%%%%%%%%%%%%%%%%%%%%%%%%%%%%%%%%%%%
\begin{abstract}
\noindent The $J=1/2$ iso-doublet double-$\Lambda$-hypernuclei, namely, ${}_{\Lambda\Lambda}^{\,\,\,\,5}$H 
and ${}_{\Lambda\Lambda}^{\,\,\,\,5}$He, are examined as the three-body cluster states, $\Lambda\Lambda t$ 
($t\equiv {}^3$H or triton) and $\Lambda\Lambda h$ ($h\equiv {}^3$He or helion), respectively, in a model 
independent framework utilizing pionless halo effective theory. Both singlet and triplet states of the
constituent $\Lambda T$ ($T\equiv t,h$) subsystem are used in the elastic channel for the study of
$_\Lambda^4$H$-\Lambda$ and $_\Lambda^4$He$-\Lambda$ scattering processes. A prototypical leading order 
investigation using a sharp momentum cutoff regulator ($\Lambda_c$) in the coupled integral equations for 
each type of the $\Lambda T$ subsystem spin states, yields identical renormalization group limit cycle behavior when the 
respective three-body contact interactions are taken close to the unitary limit. Furthermore, irrespective of the 
type of the elastic channel chosen, almost identical cutoff dependence of the three-body binding energy or 
the double-$\Lambda$-separation energy ($B_{\Lambda\Lambda}$) is obtained for the mirror partners, evidently 
suggesting good isospin symmetry in these three-body systems. Subsequently, upon normalization of our 
solutions to the integral equation with respect to a single pair of input data from an {\it ab initio} 
potential model analysis for each mirror hypernuclei, yields $B_{\Lambda\Lambda}$ which agrees fairly well 
with various erstwhile regulator independent potential models for our choice of the cutoff, 
$\Lambda_c \sim 200$~MeV. This is either consistent with pionless effective theory or with its slightly 
augmented version with a hard scale of $\Lambda_H\gtrsim 2m_\pi$, where low-energy $\Lambda$-$\Lambda$ 
interactions dominated by $\pi\pi$ or $\sigma$-meson exchange. Finally, to demonstrate the predictability of 
our effective theory, we present preliminary estimates of the $S$-wave $\Lambda\Lambda T$ three-body scattering 
lengths and the $\Lambda$-separation energies using a range of currently accepted values of the double-$\Lambda$ 
scattering length from a variety of existing phenomenological predictions that is constrained by the
recent experimental data from relativistic heavy-ion collisions. 
\end{abstract}

%\pacs{}

% 24.30.-v Resonance reactions
% 03.65.Ge Solutions of wave equations: bound states
% 14.20.-c Baryons (including antiparticles)
% 14.40.-n Mesons

%\keywords{}

\maketitle

%%%%%%%%%%%%%%%%%%%%%%%%%%%%%%%%%%%%%%%%%%%%%%%%%%%%%%%%%%%%%%%%%%%%%%%%
\section{INTRODUCTION}\label{sec:intro}
%%%%%%%%%%%%%%%%%%%%%%%%%%%%%%%%%%%%%%%%%%%%%%%%%%%%%%%%%%%%%%%%%%%%%%%%
The various experimental~\cite{Takahashi:2001nm,Ahn:2001sx,Davis:2005mb,Yoon:2007aq,Ahn:2013poa,Tamura:2013lwa,Adamczyk:2014vca,Yamamoto:2015avw,Esser:2015trs,Schulz:2016kdc,Koike:2019rrs,Acharya:2018gyz,Acharya:2019yvb} and theoretical~\cite{Jaffe:1976yi,Hammer:2001ng,Filikhin:2002wm,Filikhin:2003js,Nemura:2002hv,Myint:2002dp,Lanskoy:2003ia,Shoeb:2004cw,Nemura:2004xb,Nemura:2005ze,Ando:2013kba,Ando:2015uda,Contessi:2019csf} 
investigations over several decades on the doubly strange ($S=-2$) $s$-shell light double-$\Lambda$-hypernucler 
systems, such as, ${}_{\Lambda\Lambda}^{\,\,\,\,3}{\rm n},\,{}_{\Lambda\Lambda}^{\,\,\,\,4}{\rm n},
\,{}_{\Lambda\Lambda}^{\,\,\,\,4}{\rm H},\,{}_{\Lambda\Lambda}^{\,\,\,\,4}{\rm He},
\,{}_{\Lambda\Lambda}^{\,\,\,\,5}{\rm H},\, {}_{\Lambda\Lambda}^{\,\,\,\,5}{\rm He}$ and 
${}_{\Lambda\Lambda}^{\,\,\,\,6}{\rm He}$, have elicited keen interest in the study of exotic hypernuclei in the 
strangeness nuclear physics community. Such multistrange systems can provide stringent tests for probing the 
microscopic mechanisms for the flavor SU(3) baryon-baryon interaction in the strangeness $S=-2$ channel. In 
particular, essential information about the $\Lambda$-$\Lambda$ interaction is expected to be obtained from these studies,
which may hold definitive clues to the longstanding quest for the controversial {\it H-dibaryon}, an exotic six-quark 
($J=0,\,I=0$) deeply bound state, originally predicted by Jaffe in 1977 using the {\it bag-model}~\cite{Jaffe:1976yi}.
Different perspectives regarding the existence of the $H$ particle have been obtained in {\it ab initio} calculations
over the years. For example, the {\it dispersion relations} based analysis~\cite{Gasparyan:2011kg} on the
${}^{12}$C($K^-,K^+\Lambda\Lambda X$) reaction data from the KEK-PS Collaboration~\cite{Yoon:2007aq}, yielded an 
estimate of the ${}^1{\rm S}_0$ double-$\Lambda$ scattering length, namely, $a_{\Lambda\Lambda} =-1.2 \pm 0.6$~fm, 
that was well at odds with a possible $\Lambda\Lambda$ bound state. While lattice QCD
simulations~\cite{Beane:2010hg,Beane:2011zpa,Beane:2011iw,Inoue:2010es,Inoue:2011pg} with significantly larger 
pion masses yielded extrapolated results suggesting positive indications of a $\Lambda\Lambda$ bound state, albeit a 
shallow one in the flavor SU(3) limit. However, apparently by going to the physical point, it tends
to get pushed to the double-$\Lambda$ threshold, eventually dissolving into the continuum once SU(3) breaking 
effects are considered~\cite{Shanahan:2011su,Haidenbauer:2011ah}. In fact, of late the HAL QCD ($2+1$)-flavor coupled-channel lattice simulation~\cite{Sasaki:2019qnh} closer to the physical point 
($m^{\rm Lat}_\pi \simeq 146~{\rm MeV},\,m^{\rm Lat}_K\simeq 525~{\rm MeV}$) has yielded a rather small magnitude 
of the ${}^1{\rm S}_0$ double-$\Lambda$ scattering length, $a_{\Lambda\Lambda} =-0.81\pm 0.23$~fm, casting a 
significant doubt on the very existence of the $H$-particle. This is consistent with the current theoretically accepted
(albeit broad) range, namely, $-1.92\,\, {\rm fm} \lesssim a_{\Lambda\Lambda} \lesssim -0.5 \,\, {\rm fm}$, set by the 
fairly recent {\it thermal correlation model} based investigations~\cite{Morita:2014kza,Ohnishi:2015cnu,Ohnishi:2016elb} 
on $Au+Au$ {\it Relativistic Heavy-Ion Collisions} (RHIC) data from STAR Collaboration~\cite{Adamczyk:2014vca}, which is 
unlikely to support any $\Lambda\Lambda$ bound state. It is interesting in this regard that the same RHIC data previously 
analysed by the STAR Collaboration themselves~\cite{Adamczyk:2014vca} estimated a positive scattering length, 
$a_{\Lambda\Lambda}=1.10\pm 0.37$~fm. Nevertheless, the rather recent $\Lambda$-$\Lambda$ femtoscopic analysis of $p$-$p$ 
and $p$-Pb collision data from the ALICE Collaboration~\cite{Acharya:2018gyz,Acharya:2019yvb} yielded a $\Lambda\Lambda$
virtual bound state of energy $\approx 3.2$~MeV, thereby favoring a scattering length consistent with the above range. In 
short, although these analyses are clearly equivocal in their resolution of the $H$ particle conjecture, they evidently 
concur on a weakly attractive $\Lambda$-$\Lambda$ interaction with no deeply bound state. 

\vspace{-0.05cm}

With the discovery of ${}_{\Lambda\Lambda}^{\,\,\,\,6}$He in the hybrid-emulsion experiment 
KEK-E373~\cite{Takahashi:2001nm}, so-called the ``NAGARA'' event, along with indications of the conjectured 
${}_{\Lambda\Lambda}^{\,\,\,\,4}$H bound state in the BNL-AGS E906 production experiment~\cite{Ahn:2001sx}, arguments on the existence of double-$\Lambda$-hypernuclei have gained a firm foothold fostering a 
prolific area of modern research. A whole gamut of theoretical investigations on the 
double-$\Lambda$-hypernuclei followed since then. As for the the $J=1/2$ iso-doublet mirror partners, namely,
${}_{\Lambda\Lambda}^{\,\,\,\,5}{\rm H}$ and ${}_{\Lambda\Lambda}^{\,\,\,\,5}{\rm He}$, until rather recently most of
these investigations have been focusing on establishing phenomenological potential models. In particular, there 
exists both {\it ab initio} and {\it cluster} model approaches involving three- and four-body Faddeev-Yakubovsky 
calculations and variational 
methods~\cite{Filikhin:2002wm,Filikhin:2003js,Nemura:2002hv,Myint:2002dp,Lanskoy:2003ia,Shoeb:2004cw,Nemura:2004xb,Nemura:2005ze}. 
In some of these model analyses, the binding energy difference between the two isospin partners has been studied using 
dynamical effects of mixing between different channels, such as $\Sigma N$, $\Sigma\Sigma$, and $\Xi N$. Of these, it is 
believed that the dominant contribution arises from the $\Lambda\Lambda\,$-$\,\Xi N$ mixing channel. Because of this channel 
coupling the value of the hypernuclear binding energy (otherwise, commonly referred to in the literature as the 
{\it double-$\Lambda$-separation energy}) $B_{\Lambda\Lambda}$ of ${}_{\Lambda\Lambda}^{\,\,\,\,5}{\rm He}$ significantly 
exceeds that of ${}_{\Lambda\Lambda}^{\,\,\,\,5}{\rm H}$. However, such model approaches are often nonsystematic with 
conflicting conclusions based on {\it ad hoc} assumptions, whereby little perceptions can be gained regarding the 
underlying binding mechanisms inherent to these systems. It is, thus, timely to supplement the multitude of the existing
model results with a general model-independent prediction based on universal arguments in few-body systems.

\vspace{0.1cm}

In a recent pioneering effort, the first microscopic {\it pionless effective field theory} (${}^{\pi\!\!\!/}$EFT) 
based many-body analysis using {\it Stochastic Variational Method} (SVM) has been reported on some of the lightest 
double-$\Lambda$-hypernuclei for $A\leq 6$~\cite{Contessi:2019csf}. This kind of {\it ab initio} Hamiltonian constructed
${}^{\pi\!\!\!/}$EFT technique utilizing only elementary baryonic  ($NN,\,N\Lambda,\,\Lambda\Lambda$ two-body and 
$NNN,\,N\Lambda N,\,\Lambda N\Lambda$ three-body) interactions was first applied to calculations of few-nucleon systems
for lattice-nuclei~\cite{Barnea:2013uqa,Kirscher:2015yda,Kirscher:2017fqc} and later extended to the analysis of $s$-shell
$\Lambda$-hypernuclei~\cite{Contessi:2018qnz}. Through a {\it leading order} (LO) assessment of the onset of 
double-$\Lambda$-hypernuclei binding, the work of Ref.~\cite{Contessi:2019csf} quantitatively demonstrates the robust 
possibility of the iso-doublet partners (${}_{\Lambda\Lambda}^{\,\,\,\,5}{\rm H}\,,\,{}_{\Lambda\Lambda}^{\,\,\,\,5}{\rm He}$), 
as the lightest particle stable double-$\Lambda$-hypernuclei, thereby discounting 
${}_{\Lambda\Lambda}^{\,\,\,\,3}{\rm n}\,,\,{}_{\Lambda\Lambda}^{\,\,\,\,4}{\rm n}$ and 
${}_{\Lambda\Lambda}^{\,\,\,\,4}{\rm H}$ as possible bound states. Interestingly, as a parallel qualitative assessment to
supplement the aforementioned rigorous numerical analysis, we reexamine the 
(${}_{\Lambda\Lambda}^{\,\,\,\,5}{\rm H}\,,\,{}_{\Lambda\Lambda}^{\,\,\,\,5}{\rm He}$) iso-doublet pair in view of a 
plausible cluster or {\it halo} nuclear nature using universal arguments in physics. Particularly, in the context of 
standard ${}^{\pi\!\!\!/}$EFT framework we investigate the correlations between their bound state characteristics and the 
$S$-wave (${}_{\Lambda}^{4}{\rm H}\,$-$\,\Lambda\,,\,{}_{\Lambda}^{4}{\rm He}\,$-$\,\Lambda$) scattering processes, respectively, 
in the kinematical region below the $({}^{3}{\rm H}\,,\,{}^{3}{\rm He})+\Lambda+\Lambda$ breakup thresholds. In this way, 
through a prototypical model-independent study we assess the role of low-energy $\Lambda$-$\Lambda$ interactions in giving rise 
to universal correlations between three-body observables of such $s$-shell double-$\Lambda$-hypernuclei and their possible 
formations thereof.   

\vspace{0.1cm}

A low-energy EFT constitutes a systematic model-independent approach with low-energy observables expanded in a 
perturbative expansion in terms of a small parameter, namely, $\epsilon\sim Q/\Lambda_H \ll 1$, where $Q$ is a generic 
small momentum and $\Lambda_H$ is the ultraviolet (UV) cutoff scale which limits the applicability of the perturbative 
scheme. The effective degrees of freedom consistent with the low-energy symmetries of the system are then identified in
terms of which the Lagrangian of the system is constructed and expanded in increasing order of derivative interaction. The corresponding coefficients (low-energy constants) are fixed from phenomenological data. The heavy degrees of freedom
above the hard scale $\Lambda_H$ are integrated out and their effects are implicitly encoded in these couplings. In the 
so-called {\it halo/cluster} EFT formalism, the ${}_{\Lambda\Lambda}^{\,\,\,\,5}{\rm H}$ and 
${}_{\Lambda\Lambda}^{\,\,\,\,5}{\rm He}$ systems can be regarded as the double-$\Lambda$ {\it halo\,}-nuclear states, namely, 
$\Lambda\Lambda t$ ($t\equiv {}^3$H, i.e., the {\it triton}) and $\Lambda\Lambda h$ ($h\equiv {}^3$He, i.e., the 
{\it helion}), respectively, with $T\equiv t,h$ being the compact core that can be considered elementary at scales chosen 
well below the breakup of ${}_{\Lambda}^{4}{\rm H}$ and ${}_{\Lambda}^{4}{\rm He}$.   
 
\vspace{0.1cm}
 
The erstwhile emulsion works~\cite{Juric:1973zq,Davis:2005mb,Tamura:2013lwa} have indicated evidences of particle 
stable states of ${}_{\Lambda}^{4}{\rm H}$ and ${}_{\Lambda}^{4}{\rm He}$ $\Lambda$-hypernuclei. The existence of 
these states were recently reconfirmed by high-resolution decay $\pi^-$ and $\gamma$-ray spectroscopic measurements 
carried out by the A1 Collaboration at MAMI~\cite{Esser:2015trs,Schulz:2016kdc} and the E13 Collaboration at 
J-PARC~\cite{Yamamoto:2015avw,Koike:2019rrs}, respectively. The extracted $J^p=0^+$ ground state 
$\Lambda$-{\it separation energies} (${\mathcal B}_\Lambda[0^+]$) of ${}_\Lambda^4$H and ${}_\Lambda^4$He are 
$2.157\pm 0.077$~MeV and $2.39\pm 0.05$~MeV, respectively, whereas those corresponding to the $J^p=1^+$ first excited 
state (${\mathcal B}_\Lambda[1^+]$) are $1.067\pm0.08$~MeV and $0.984\pm 0.05$~MeV, respectively (cf. level scheme 
depicted in Fig.~\ref{fig:fig0}). Thus, the typical momentum scale $Q$ associated with these single 
$\Lambda$-hypernuclei can be naively identified with mean binding momentum of the ground and first excited states, 
namely, $\bar{Q}\sim\sqrt{\mu_{\Lambda T} \left({\mathcal B}_{\Lambda}[0^+]+{\mathcal B}_{\Lambda}[1^+]\right)}\approx 50$~MeV,  
with $\mu_{\Lambda T}=M_\Lambda M_T/(M_\Lambda +M_T)$ being the reduced mass of these $\Lambda T$ subsystems. On the 
other hand, the experimental binding energies (${\mathcal B}_T$) of the triton and helion cores being $8.48$~MeV and 
$7.72$~MeV, respectively, the {\it breakdown} scale of our EFT framework may be associated with the corresponding 
binding momentum scale $\Lambda_H\sim\sqrt{2\mu_{d N} {\mathcal B_T}}\sim m_\pi$ of the cores, with $\mu_{dN}$ being the
reduced mass of the deuteron ($d$) and nucleon ($N$) system, and $m_\pi$ is the pion mass. Consequently, the expansion
parameter is conservatively estimated to be at the most 
$\epsilon \sim \bar{Q}/m_\pi \lesssim \sqrt{2\mu_{\Lambda T} {\mathcal B}_{\Lambda}[0^+]}/m_\pi \approx 0.4$, a value 
reasonably small to support a valid EFT framework. 

\vspace{0.1cm}

A practical computational framework for investigating three-body dynamics is thus provided by the ${}^{\pi\!\!\!/}$EFT 
without explicit inclusion of pion. This has become a popular tool for investigating shallow bound state systems of 
nucleons and other hadrons (for reviews and relatively recent works, e.g., see
Refs.~\cite{Hammer:2001ng,Kaplan:1998tg,Kaplan:1998we,vanKolck:1998bw,Bedaque:1998kg,Bedaque:1998km,Bedaque:1999ve,Braaten:2004rn,Ando:2013kba,Ando:2015uda,Ando:2015fsa,Raha:2017ahu} 
and other references therein). Such a framework provides the most general 
approach to handle the dynamics of finely tuned systems with large scattering lengths and cross sections nearly saturating 
the unitary bound. This happens presumably in the vicinity of nontrivial renormalization group (RG) fixed points of the 
two-body contact couplings. Recently, a large number of works on ${}^{\pi\!\!\!/}$EFT have appeared dealing with low-energy
universal physics of three-body systems. A typical signature of the onset of such universality is the appearance of a RG 
limit cycle resulting from the breakdown of an exact to a discrete scaling symmetry, accompanied with the emergence of a 
geometric tower of arbitrary shallow three-body Efimov bound states~\cite{Efimov:1970zz,Braaten:2004rn}. In the context of
hypernuclear physics, the Efimov effect and its universal role in the prediction of three-body exotic bound
states have been discussed in a number of theoretical
works~\cite{Hammer:2001ng,Ando:2013kba,Ando:2015uda,Ando:2015fsa,Hildenbrand:2019sgp} based on ${}^{\pi\!\!\!/}$EFT at LO.
In the ensuing analysis, we use a similar set-up to investigate whether any remnant universal feature inherent to the 
$\Lambda\Lambda T$ system indicates Efimov-like bound state character. However, the current paucity of phenomenological 
information to constrain the various low-energy parameters of the theory is a major hurdle in our approach which 
precludes a robust prediction of the existence of Efimov-like bound states in the ${}_{\Lambda\Lambda}^{\,\,\,\,5}{\rm H}$ 
and ${}_{\Lambda\Lambda}^{\,\,\,\,5}{\rm He}$ systems. As demonstrated in our analysis, a crucial piece of information 
required as input to the EFT analysis is a three-body datum, namely, the three-body binding or double-$\Lambda$-separation
energy $B_{\Lambda\Lambda}$ of a given mirror partner, for which there are currently no available experimental estimates. For
this purpose, we rely on suitable predictions based on an existing potential models, e.g., the {\it ab initio} SVM analysis 
of Nemura {\it et al.}~\cite{Nemura:2004xb}. Moreover, the predictability of our halo/cluster EFT framework depends on 
fixing several two-body parameters from the following phenomenological information: 
%--figure---------------------------------
\begin{figure}[tbp]
    \centering
    \includegraphics[scale=0.54]{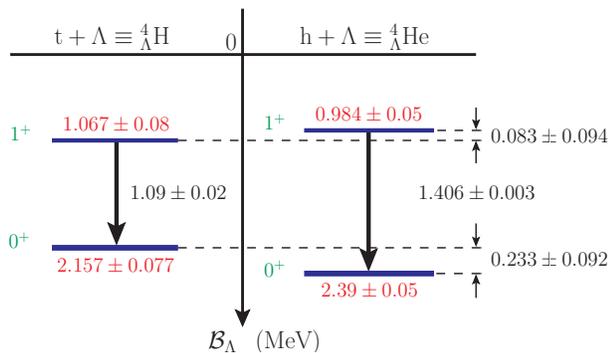}
    \caption{\label{fig:fig0} Level energy (${\mathcal B}_\Lambda$) scheme with the ground ($J^P=0^+$) state of 
             ${}_{\Lambda}^{4}{\rm H}$ and the first-excited ($J^P=1^+$) states of the mirror partners 
             (${}_{\Lambda}^{4}{\rm H},\,{}_{\Lambda}^{4}{\rm He}$) taken from the recent high-resolution spectroscopic 
             measurements at MAMI~\cite{Esser:2015trs,Schulz:2016kdc} and J-PARC~\cite{Yamamoto:2015avw,Koike:2019rrs},
             respectively. The ground state energy of ${}_{\Lambda}^{4}{\rm He}$ on the other hand is taken from the 
             erstwhile emulsion work of Ref.~\cite{Davis:2005mb}. The figure is adapted from 
             Refs.~\cite{Schulz:2016kdc,Gazda:2016qva}.}
\end{figure}
%--figure---------------------------------
\begin{itemize}
\item the measured ground and first-excited state $\Lambda$-separation energies ${\mathcal B}_\Lambda[J^P=0^+,1^+]$ 
of the mirror $\Lambda$-hypernuclei (${}_{\Lambda}^{4}{\rm H}\,,\,{}_{\Lambda}^{4}{\rm He}$), which we take from
Refs.~\cite{Davis:2005mb,Yamamoto:2015avw,Esser:2015trs,Schulz:2016kdc,Koike:2019rrs} (cf. Fig.~\ref{fig:fig0}); and 
\item second, the value of the $S$-wave double-$\Lambda$ scattering length $a_{\Lambda\Lambda}$, for which we consider an
acceptable range of values from various phenomenological analyses~\cite{Contessi:2019csf,Gasparyan:2011kg,Sasaki:2019qnh,Morita:2014kza,Ohnishi:2015cnu,Ohnishi:2016elb,Nagels:1978sc,Rijken:1998yy,Stoks:1999bz}, 
constrained by the recent RHIC data~\cite{Adamczyk:2014vca}.
\end{itemize}
Based on these inputs, the three-body integral equations completely determine the $B_{\Lambda\Lambda}\,$-$\,a_{\Lambda\Lambda}$ 
correlations for the $\Lambda\Lambda T$ systems, using which preliminary estimates of the corresponding $S$-wave three-body 
scattering lengths $a_{\Lambda\Lambda T}$ are predicted. Such EFT predicted scattering lengths induce universal 
correlations between three-body observables, as elucidated by the so-called {\it Phillips-lines}~\cite{Phillips:1968zze} 
(cf. Fig.~\ref{fig:fig9}). Furthermore, for the recently suggested benchmark value, $a_{\Lambda\Lambda}=-0.80$ fm, in 
Ref.~\cite{Contessi:2019csf}, the $\Lambda$-separation energies,
${\mathcal B}_{\Lambda}({}_{\Lambda\Lambda}^{\,\,\,\,5}{\rm H})=2.295$ MeV and 
${\mathcal B}_{\Lambda}({}_{\Lambda\Lambda}^{\,\,\,\,5}{\rm He})=2.212$ MeV, are deduced.
    
\vspace{0.1cm}
    
The paper is organized as follows. In Sec.~\ref{sec:theory} we present the basic set-up of the ${}^{\pi\!\!\!/}$EFT formalism. 
There we display the most general LO effective Lagrangian and the coupled system of integral equations for the 
$\Lambda\Lambda T$ system, with appropriate scale dependent three-body contact interactions that describe RG limit cycle 
behavior. Section ~\ref{sec:results} contains our numerical results of solving the integral equations in both bound and scattering
domains. In particular, through our study of the $B_{\Lambda\Lambda}\,$-$\,a_{\Lambda\Lambda}$ correlations, we present preliminary
estimates of the $\Lambda\Lambda T$ scattering lengths and the corresponding $\Lambda$-separation energies. Finally, in 
Sec.~\ref{sec:summary} we present our summary with concluding remarks. A brief discussion on the one- and two-body 
non-relativistic propagators in ${}^{\pi\!\!\!/}$EFT is relegated to the appendix.  

\vspace{-0.2cm}

%%%%%%%%%%%%%%%%%%%%%%%%%%%%%%%%%%%%%%%%%%%%%%%%%%%%%%%%%%%%%%%%%%%%%%%%
\section{THEORETICAL FRAMEWORK}\label{sec:theory}
%%%%%%%%%%%%%%%%%%%%%%%%%%%%%%%%%%%%%%%%%%%%%%%%%%%%%%%%%%%%%%%%%%%%%%%%
%%%%%%%%%%%%%%%%%%%%%%%%%%%%%%%%%
\subsection{Effective Lagrangian}
%%%%%%%%%%%%%%%%%%%%%%%%%%%%%%%%%
We use the theoretical framework of pionless effective field theory to investigate the bound states of the 
double-$\Lambda$-hypernuclear mirror systems (${}_{\Lambda\Lambda}^{\,\,\,\,5}$H\,, ${}_{\Lambda\Lambda}^{\,\,\,\,5}$He). 
In this approach the effective Lagrangian is constructed manifestly nonrelativistic on the basis of available 
symmetries of the relevant low-energy degrees of freedom. In our case, the explicit elementary degrees of freedom 
involve two $\Lambda$-hyperon {\it halo} fields and the generic {\it core} field, $T\equiv t,h$, representing one of
the two mirrors (isospin) partners, namely, the triton ($t$) or the helion ($h$). In addition, it is convenient to 
introduce auxiliary dimer fields to unitarize and renormalize the two-body 
sectors~\cite{Bedaque:1998km,Braaten:2004rn,BS01,AH04,Ando:2010wq}. Our formalism includes three such dimer fields, 
namely, the spin-singlet (${}^{1}{\rm S}_0$) field $u_0\equiv (\Lambda T)_{s}$, the spin-triplet (${}^{3}{\rm S}_1$) 
field $u_1\equiv (\Lambda T)_{t}$, and the spin-singlet $\Lambda\Lambda$-dibaryon field $u_s\equiv (\Lambda\Lambda)_s$. 
Notably, these $u_{0}$ and $u_{1}$ dimer states correspond to the experimentally observed spin-singlet ($0^+$) ground state 
and spin-triplet ($1^+$) excited state of the mirror hypernuclei $({}_{\Lambda}^{4}{\rm H}\,,\,{}_{\Lambda}^{4}{\rm He})$~\cite{Juric:1973zq,Davis:2005mb,Tamura:2013lwa,Esser:2015trs,Schulz:2016kdc,Yamamoto:2015avw,Koike:2019rrs}.

\vspace{0.1cm}

The full nonrelativistic LO ${}^{\pi\!\!\!/}$EFT Lagrangian can be expressed as the following string of terms: 
\begin{equation}
\mathcal{L} = \mathcal{L}_\Lambda + \mathcal{L}_T + \mathcal{L}_{u_0} + \mathcal{L}_{u_1} + \mathcal{L}_{u_s} 
+ \mathcal{L}_{\rm 3{\text -}body}\,.
\end{equation}
The one-body Lagrangian containing the contributions of the elementary fields, namely, the $\Lambda$-hyperon 
and the spin-1/2 core $T$, is given as  
\begin{eqnarray}
\mathcal{L}_\Lambda&=&\Lambda^{\dagger}\bigg[i(v\cdot\partial)+\frac{(v\cdot\partial)^2
-\partial^2}{2M_\Lambda}+\cdots\bigg]\Lambda\,\,,
\\
\mathcal{L}_T&=&T^{\dagger}\bigg[i(v\cdot\partial)+\frac{(v\cdot\partial)^2-\partial^2}{2M_T}+\cdots\bigg]T\,\,,
\end{eqnarray}
where $M_\Lambda$ and $M_T$ are the respective masses of the elementary fields. Next we display the two-body parts of 
the Lagrangian, namely,
\begin{eqnarray}
\mathcal{L}_{u_0} &=&-u_0^{\dagger}\bigg[i(v\cdot\partial)+\frac{(v\cdot\partial)^2-\partial^2}{2(M_\Lambda+M_T)}+\cdots\bigg]u_0
\nonumber \\
&&-\,y_0\bigg[u_0^{\dagger}\left(T^{\rm T}\,\hat{\mathbb P}^{({}^{1}{\rm S}_0)}_{(\Lambda T)}\,\Lambda\right)+ \rm{h.c.}\bigg]+\cdots\,\,,
\\ 
\nonumber\\
\mathcal{L}_{u_1} &=&-(u_1)_j^{\dagger}\bigg[i(v\cdot\partial)+\frac{(v\cdot\partial)^2-\partial^2}{2(M_\Lambda+M_T)}+\cdots\bigg](u_1)_j
\nonumber \\
&&-\,y_1\bigg[(u_1)_j^{\dagger}\left(T^{\rm T}\,\hat{\mathbb P}_{(\Lambda T)\,j}^{({}^{3}{\rm S}_1)}\,\Lambda\right)+ \rm{h.c.}\bigg]+\cdots\,\,,
\\
\nonumber\\
\mathcal{L}_{u_s} &=&-u_s^{\dagger}\bigg[i(v\cdot\partial)+\frac{(v\cdot\partial)^2-\partial^2}{4M_\Lambda}+\cdots\bigg]u_s
\nonumber \\
&&-\,y_s\bigg[u_s^{\dagger}\left(\Lambda^{\rm T}\,\hat{\mathbb P}^{({}^{1}{\rm S}_0)}_{(\Lambda\Lambda)}\,\Lambda\right)+ \rm{h.c.}\bigg]+\cdots\,\,,
\end{eqnarray}   
where the spin-singlet and spin-triplet projection operators are given as
\begin{eqnarray}
\hat{\mathbb P}_{(\Lambda \Lambda)}^{({}^{1}{\rm S}_0)}=-\frac{i}{2}\sigma_2 \,,\quad && \quad
\hat{\mathbb P}_{(\Lambda T)}^{({}^{1}{\rm S}_0)}=-\frac{i}{\sqrt{2}}\sigma_2 \,, 
\nonumber\\
\hat{\mathbb P}_{(\Lambda T)\,j}^{({}^{3}{\rm S}_1)}&=&-\frac{i}{\sqrt{2}}\sigma_2\sigma_j\,,
\end{eqnarray}
with $\sigma_j\,(j=1,2,3)$ being the Pauli spin matrices. In the above equations $v^{\mu}=(1,{\bf 0})$ 
is the velocity four-vector, and the couplings $y_{0}$, $y_{1}$, and $y_{s}$ are two-body contact interactions between the 
respective dimer and their constituent elementary fields. Adopting to the power-counting scheme for the 
contact interactions apposite to finely tuned systems~\cite{Kaplan:1998tg,Kaplan:1998we,vanKolck:1998bw}, 
these LO couplings are easily fixed as~\cite{Griesshammer:2004pe} 
\begin{eqnarray}
 y_0 = &y_1& = \sqrt{\frac{2\pi}{\mu_{\Lambda T}}}\,,
 \nonumber\\
 \nonumber\\
 \text{and} \qquad y_s &=&\sqrt{\frac{4\pi}{M_\Lambda}}\,.
\label{eq:y01s}
\end{eqnarray}
The ellipses in all the above Lagrangians denote subleading order terms containing four or higher derivative
operators that do not contribute to our LO analysis. For pedagogical reasons a brief description of the 
one- and two-body nonrelativistic propagators used in the construction of the Faddeev-type coupled integral 
equations is presented in the appendix. 

\vspace{0.1cm}

Finally, as demonstrated later in this section, since the $\Lambda\Lambda T$ three-body systems are found to 
exhibit RG limit cycle behavior, the set of integral equations [cf. Eqs.~\eqref{eq:type-A} and \eqref{eq:type-B}\,] 
becomes ill-defined in the asymptotic UV limit, and a regulator, say, in the form of a sharp momentum cutoff 
$\Lambda_c$ must be introduced to obtain regularized finite results. In that case, the basic tenet of the 
EFT~\cite{Bedaque:1998km} demands the introduction of nonderivatively coupled LO counterterms to renormalize 
the {\it artificial} regulator ($\Lambda_c$) dependence of the integral equations. For the $\Lambda\Lambda T$ 
($J=1/2,I=1/2$) mirror systems, there exists two equivalent choices for the subsystem spin rearrangements that 
determine the elastic channels, namely, $u_0\Lambda\to u_0\Lambda$ (denoted ``type-A''), and 
$u_1\Lambda\to u_1\Lambda$ (denoted``type-B''). With the type-A, and -B choices of the elastic channels, the 
three-body counterterm Lagrangians are 
\begin{widetext}

\vspace{-0.2cm}

\begin{eqnarray}
\mathcal{L}^{(A)}_{\rm 3{\text -}body}&=&-\frac{g^{(A)}_3(\Lambda_c)}{\Lambda_c^2}\left[-\frac{M_T y_0^2}{2}(u_0\Lambda)^{\dagger}(u_0\Lambda)+\frac{M_Ty_0y_1}{2}(u_0\Lambda)^{\dagger}\left({\bf u}_1\cdot\boldsymbol{\sigma}\Lambda\right)-\frac{M_{\Lambda}y_sy_0}{\sqrt{2}}(u_0\Lambda)^{\dagger}(u_sT)+ {\rm h.c.}\right]\,, 
\\
\nonumber\\
\mathcal{L}^{(B)}_{\rm 3{\text -}body}&=&-\frac{g_3^{(B)}(\Lambda_c)}{\Lambda_c^2}\left[\frac{M_T y_1^2}{6} \left({\bf u}_1\cdot\boldsymbol{\sigma}\Lambda\right)^{\dagger}\left({\bf u}_1\cdot\boldsymbol{\sigma}\Lambda\right)+\frac{M_T y_0y_1}{2} \left({\bf u}_1\cdot\boldsymbol{\sigma}\Lambda\right)^{\dagger}(u_0\Lambda)-\frac{M_{\Lambda} y_sy_1}{\sqrt{2}} \left({\bf u}_1\cdot\boldsymbol{\sigma}\Lambda\right)^{\dagger}(u_sT)+ {\rm h.c.}\right].\qquad
\label{eq:type_3body}
\end{eqnarray}
\end{widetext}
The regulator dependent three-body running couplings $g^{(A)}_3(\Lambda_c)$ and $g^{(B)}_3(\Lambda_c)$ which are used to absorb the 
scale dependence of the integral equations are {\em a priori} undetermined in the EFT. Hence they must be 
phenomenologically fixed from essential three-body data. A typical signature that Efimov 
physics~\cite{Efimov:1970zz,Braaten:2004rn} is manifest in the three-body system is that the RG behavior of the 
three-body couplings $g^{(A)}_3$ and $g^{(B)}_3$ displays a characteristic quasi-log cyclic periodicity as a function of the 
regulator scale $\Lambda_c\ll \infty$. As originally suggested by Wilson~\cite{Wilson}, this unambiguously implies
the onset of an RG limit cycle. Here we note that exact universality demands both three-body couplings to be 
identical which in principle should not depend on the details of the two-body subsystems. However, in practice,
certain nominal qualitative differences do appear in the estimation of these scale dependent couplings, as seen in 
our results presented in the next section. This is primarily due to the specific choice of the renormalization 
schemes we have adopted in the treatments of the type-A and type-B integral equations [cf. discussion below 
Eq.~\eqref{eq:g3_modified_K}]. However, such differences do not have any significant influence on the qualitative 
nature of the results of this work. 

%--figure---------------------------------
\begin{figure*}[tbp]
    \centering
    \includegraphics[width=17.5cm]{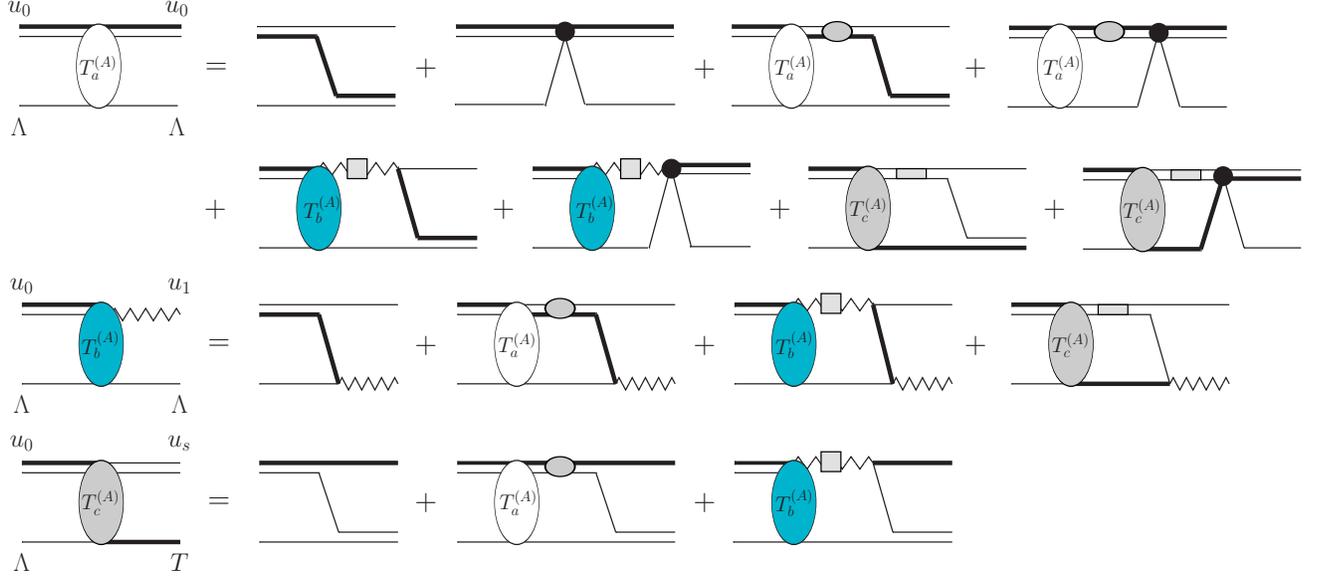}
    \caption{\label{fig:fig1} Feynman diagrams for the coupled-channel integral equations, with 
             $u_0\Lambda\to u_0\Lambda$ (type-A) choice as the elastic channel. The thin (thick) 
             lines denote the $\Lambda$-hyperon (core $T\equiv t,h$) field propagators. The 
             double lines denote the renormalized propagators for the spin-singlet dimer fields $u_{0}$ 
             and $u_{s}$, and the zigzag lines denote the renormalized propagators for the spin-triplet 
             dimer field $u_1$. The dark filled circles denote the leading order three-body contact 
             interactions, while the square, oval, and rectangular gray blobs represent dressings 
             of the dimer propagators with resummed loops (cf. discussion in the appendix).}
\end{figure*}
%--figure---------------------------------

%\vspace{-0.2cm}

%%%%%%%%%%%%%%%%%%%%%%%%%%%%%%%
\subsection{Integral equations}
\begin{figure*}[tbp]
    \centering
    \includegraphics[width=17.5cm]{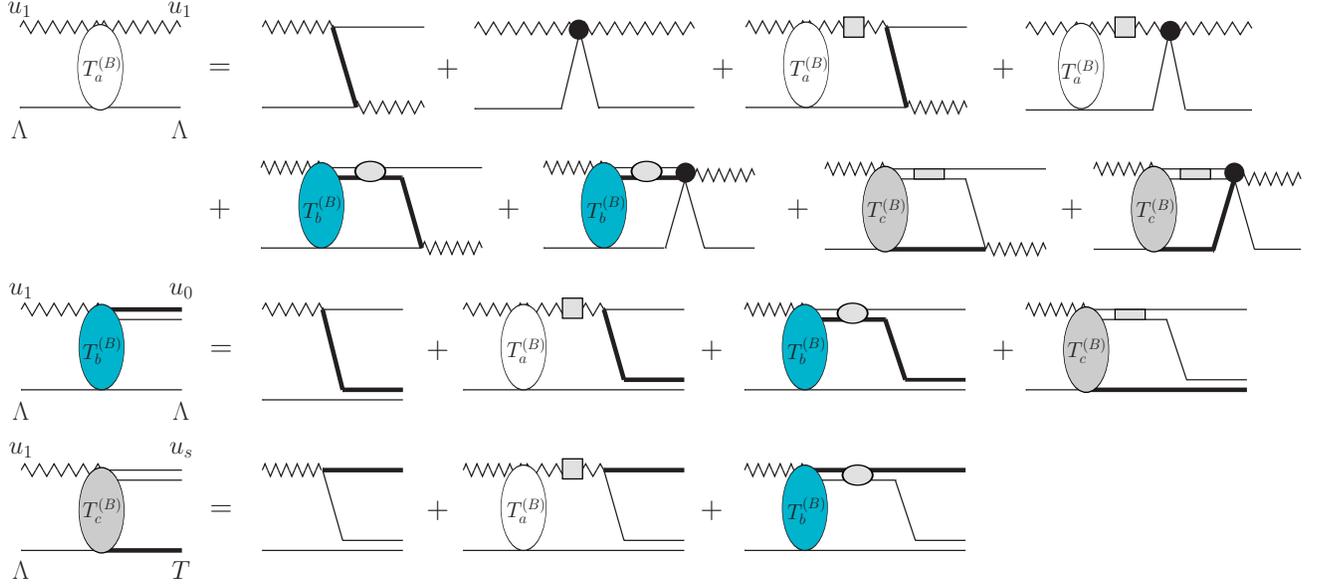}
    \caption{\label{fig:fig2}  Feynman diagrams for the coupled-channel integral equations, with 
             $u_1\Lambda\to u_1\Lambda$ (type-B) choice for the elastic channel. The thin (thick) 
             lines denote the $\Lambda$-hyperon (core $T\equiv t,h$) field propagators. The 
             double lines denote the renormalized propagators for the spin-singlet dimer fields $u_{0}$
             and $u_{s}$, and the zigzag lines denote the renormalized propagators for the spin-triplet 
             dimer field $u_1$. The dark filled circles denote the leading order three-body contact 
             interactions, while the square, oval, and rectangular gray blobs represent dressings 
             of the dimer propagators with resummed loops (cf. discussion in the appendix). }
\end{figure*}

%%%%%%%%%%%%%%%%%%%%%%%%%%%%%%%
In Figs.~\ref{fig:fig1} and \ref{fig:fig2}, we display the Feynman diagrams contributing to the $S$-wave elastic 
processes, namely, $u_0\Lambda \rightarrow u_0\Lambda$ (type-A) and $u_1\Lambda \rightarrow  u_1\Lambda$  (type-B),
in terms of the {\it half-off-shell} $S$-wave projected amplitudes, 
$T^{(A,B)}_a(p,k;E),\,T^{(A,B)}_b(p,k;E)$ and $T^{(A,B)}_c(p,k;E)$. While $T^{(A,B)}_a(p,k;E)$ denotes the elastic 
amplitudes, $T^{(A,B)}_b(p,k;E)$ and $T^{(A,B)}_c(p,k;E)$ are the amplitudes for the inelastic processes, 
$u_{0,1}\Lambda\to u_{1,0}\Lambda$ and $u_{0,1}\Lambda\to u_{s}\Lambda$, respectively. Here $k$ ($p$) is the 
relative on-shell (off-shell) three-body center-of-mass momentum for the $u_{0,1}\,$-$\,\Lambda$ scattering processes 
in the initial (final) states, and $E={\mathcal E}^{thr}_{2(s,t)}+k^2/(2\mu_{\Lambda(\Lambda T)})$ is the total 
center-of-mass kinetic energy measured with respect to the {\it three-particle breakup} threshold $(E=0)$. In other
words, for each $\Lambda\Lambda T$ three-body system, there exists two {\it particle-dimer breakup} thresholds, viz. 
the deeper $\Lambda+u_0$ breakup threshold, ${\mathcal E}^{thr}_{2(s)}=-\gamma^2_0/(2\mu_{\Lambda T})$, and the 
shallower $\Lambda+u_1$ breakup threshold, ${\mathcal E}^{thr}_{2(t)}=-\gamma^2_1/(2\mu_{\Lambda T})$ (cf. 
discussions in Sec.~\ref{sec:results}\,). Here $\gamma_{0}$ and $\gamma_{1}$ are the respective binding momenta of 
the singlet $u_0\equiv (\Lambda T)_s$ and triplet $u_1\equiv (\Lambda T)_t$ two-body subsystems, and 
$\mu_{\Lambda(\Lambda T)}=M_\Lambda(M_\Lambda+M_T)/(2 M_\Lambda +M_T)$ is the reduced masses of the 
$\Lambda\,$-$\,(\Lambda T)_{s,t}$ three-body system. Using standard Feynman rules, the $S$-wave projected amplitudes for 
the different elastic and inelastic channels can be easily worked out. With the type-A and type-B choices of the elastic 
channels, the two sets of coupled integral equations for the $\Lambda\Lambda T$ mirror partners are given 
as~\cite{STM1,STM2,DL61,DL63}
\begin{widetext}
%%%%%%%%%%%%%%%%%%%%%%%
\begin{eqnarray}
T^{(A)}_a(p,k;E) 
&=& -\frac{1}{2}(y_0^2M_T){\mathcal K}^{A}_{(a)}(p,k;E)
+\frac{M_T}{\mu_{\Lambda T}}\int_0^{\Lambda_c} \frac{dq\,q^2}{2\pi} {\mathcal K}^{A}_{(a)}(p,q,\Lambda_c;E)\,{\mathcal D}_{0}(q,E)\,{T^{(A)}_a(q,k;E)}
\nonumber\\
\nonumber\\
&&\hspace{-1.5cm}-\,\frac{y_0}{y_1}\frac{\sqrt{3}M_T}{\mu_{\Lambda T}}\int_0^{\Lambda_c} \frac{dq\,q^2}{2\pi} {\mathcal K}^{A}_{(a)}(E;p,q)\,{\mathcal D}_{1}(q,E)\,{T^{(A)}_b(q,k;E)}
+\frac{y_0}{y_s}\sqrt{8}\int_0^{\Lambda_c} \frac{dq\,q^2}{2\pi} {\mathcal K}^{A}_{(b2)}(p,q;E)\,{\mathcal D}_{s}(q,E)\,{T^{(A)}_c(q,k;E)}\,,
\nonumber\\
\nonumber\\
%%%%%%%%%%%%%%%%%%%%%%%
T^{(A)}_b(p,k;E) 
&=& \frac{\sqrt{3}}{2}(y_0y_1M_T)K_{(a)}(p,k;E)
-\frac{y_1}{y_0}\frac{\sqrt{3}M_T}{\mu_{\Lambda T}}\int_0^{\Lambda_c} \frac{dq\,q^2}{2\pi} K_{(a)}(p,q;E)\,{\mathcal D}_{0}(q,E)\,{T^{(A)}_a(q,k;E)}
\nonumber\\
\nonumber\\
&&\hspace{-1.5cm}-\,\frac{M_T}{\mu_{\Lambda T}}\int_0^{\Lambda_c} \frac{dq\,q^2}{2\pi} K_{(a)}(p,q;E)\,{\mathcal D}_{1}(q,E)\,{T^{(A)}_b(q,k;E)}
+\frac{y_1}{y_s}\sqrt{24}\int_0^{\Lambda_c} \frac{dq\,q^2}{2\pi} K_{(b2)}(p,q;E)\,{\mathcal D}_{s}(q,E)\,{T^{(A)}_c(q,k;E)}\,,
\nonumber \\
\nonumber \\
%%%%%%%%%%%%%%%%%%%%%%%
T^{(A)}_c(p,k;E) 
&=& -\frac{1}{\sqrt{2}}{(y_0y_sM_\Lambda)}K_{(b1)}(p,k;E)
+\frac{y_s}{y_0}\frac{\sqrt{2}M_\Lambda}{\mu_{\Lambda T}}\int_0^{\Lambda_c} \frac{dq\,q^2}{2\pi} K_{(b1)}(p,q;E)\,{\mathcal D}_{0}(q,E)\,{T^{(A)}_a(q,k;E)}
\nonumber\\
\nonumber\\
&&\hspace{-1.5cm}+\frac{y_s}{y_1}\,\frac{\sqrt{6}M_\Lambda}{\mu_{\Lambda T}}\int_0^{\Lambda_c} \frac{dq\, q^2}{2\pi} K_{(b1)}(p,q;E)\,{\mathcal D}_{1}(q,E)\,{T^{(A)}_b(q,k;E)}\,,
\label{eq:type-A}
\end{eqnarray}
%%%%%%%%%%%%%%%%%%%%%%%
and, 
%%%%%%%%%%%%%%%%%%%%%%%
\begin{eqnarray}
T^{(B)}_a(p,k;E) 
&=& \frac{1}{2}(y_1^2M_T)\mathcal{K}^{B}_{(a)}(p,k;E) 
-\frac{M_T}{\mu_{\Lambda T}}\int_0^{\Lambda_c} \frac{dq\,q^2}{2\pi} {\mathcal K}^{B}_{(a)}(p,q,\Lambda_c;E)\,{\mathcal D}_{1}(q,E)\,{T^{(B)}_a(q,k;E)}
\nonumber\\
\nonumber\\
&&\hspace{-1.5cm}-\,\frac{y_1}{y_0}\frac{\sqrt{3}M_T}{\mu_{\Lambda T}}\int_0^{\Lambda_c} \frac{dq\,q^2}{2\pi} {\mathcal K}^{B}_{(a)}(p,q;E)\,{\mathcal D}_{0}(q,E)\,{T^{(B)}_b(q,k;E)}
+\frac{y_1}{y_s}\sqrt{24}\int_0^{\Lambda_c} \frac{dq\,q^2}{2\pi} {\mathcal K}^{B}_{(b2)}(p,q;E)\,{\mathcal D}_{s}(q,E)\,{T^{(B)}_c(q,k;E)}\,\,,
\nonumber\\
\nonumber\\
%%%%%%%%%%%%%%%%%%%%%%%
T^{(B)}_b(p,k;E) 
&=&  \frac{\sqrt{3}}{2}(y_1y_0M_T)K_{(a)}(p,k;E)
-\frac{y_0}{y_1}\frac{\sqrt{3}M_T}{\mu_{\Lambda T}}\int_0^{\Lambda_c} \frac{dq\,q^2}{2\pi} K_{(a)}(p,q;E)\,{\mathcal D}_{1}(q,E)\,{T^{(B)}_a(q,k;E)}
\nonumber\\
\nonumber\\
&&\hspace{-1.5cm}+\,\frac{M_T}{\mu_{\Lambda T}}\int_0^{\Lambda_c} \frac{dq\,q^2}{2\pi} K_{(a)}(p,q;E)\,{\mathcal D}_{0}(q,E)\,{T^{(B)}_b(q,k;E)}
+\frac{y_0}{y_s}\sqrt{8}\int_0^{\Lambda_c} \frac{dq\,q^2}{2\pi} K_{(b2)}(p,q;E)\,{\mathcal D}_{s}(q,E)\,{T^{(B)}_c(q,k;E)}\,\,,
\nonumber\\
\nonumber\\
%%%%%%%%%%%%%%%%%%%%%%%
T^{(B)}_c(p,k;E) 
&=& -\sqrt{\frac{3}{2}}{(y_1y_sM_\Lambda)}K_{(b1)}(p,k;E)
+\,\frac{y_s}{y_1}\frac{\sqrt{6}M_\Lambda}{\mu_{\Lambda T}}\int_0^{\Lambda_c} \frac{dq\,q^2}{2\pi} K_{(b1)}(p,q;E)\,{\mathcal D}_{1}(q,E)\,{T^{(B)}_a(q,k;E)}
\nonumber\\
\nonumber\\
&&\hspace{-1.5cm}+\,\frac{y_s}{y_0}\frac{\sqrt{2}M_\Lambda}{\mu_{\Lambda T}}\int_0^{\Lambda_c} \frac{dq\,q^2}{2\pi} K_{(b1)}(p,q;E)\,{\mathcal D}_{0}(q,E)\,{T^{(B)}_b(q,k;E)}\,,
\label{eq:type-B}
\end{eqnarray}
%%%%%%%%%%%%%%%%%%%%%%%
\end{widetext}
respectively, where in the above equations the two-body couplings $y_{0}$, $y_{1}$, and $y_{s}$ are determined by using Eq.~\eqref{eq:y01s}. 
The $S$-wave projected two-point Green's functions (cf. Eq.~\eqref{dimer_props} in the appendix), namely,
%--figure---------------------------------
%--figure---------------------------------
\begin{eqnarray}
 \mathcal{D}_{0}(q,E) &=& \frac{1}{\gamma_0 -\sqrt{q^2\frac{\mu_{\Lambda T}}{\mu_{\Lambda(\Lambda T)}}-2\mu_{\Lambda T} E-i\eta}-i\eta}\,,
\nonumber\\ 
\nonumber\\ 
 \mathcal{D}_{1}(q,E) &=& \frac{1}{\gamma_1 -\sqrt{q^2\frac{\mu_{\Lambda T}}{\mu_{\Lambda(\Lambda T)}}-2\mu_{\Lambda T} E-i\eta}-i\eta}\,,
\nonumber \\
\nonumber\\ 
 \mathcal{D}_{s}(q,E) &=& \frac{1}{\frac{1}{a_{\Lambda\Lambda}}-\sqrt{q^2\frac{M_\Lambda}{2\mu_{T(\Lambda\Lambda )}}-M_\Lambda E-i\eta}-i\eta},\,\quad
\end{eqnarray}
contain the contributions of the $u_{0}$, $u_{1}$, and $u_{s}$ intermediate dimer states, with 
$\mu_{T(\Lambda\Lambda )}=(2M_\Lambda M_T)/(2 M_\Lambda +M_T)$, which is the reduced mass of the $T\,$-$\,(\Lambda\Lambda)_s$ 
three-body system. The $T$-exchange interaction kernel $K_{(a)}$, and the two possible $\Lambda$-exchange 
interaction kernels, $K_{(b1)}$ and $K_{(b2)}$, can be expressed as
\begin{eqnarray}
K_{(a)}(p,\kappa;E) \!&=&\! \frac{1}{2p \kappa}\ln\bigg[\frac{p^{2}+\kappa^2+\frac{2\mu_{\Lambda T}}{M_T}p \kappa-2\mu_{\Lambda T} E}{p^{2}+\kappa^2-\frac{2\mu_{\Lambda T}}{M_T}p \kappa-2\mu_{\Lambda T} E}\bigg]\,,
\nonumber
\end{eqnarray}
and
\begin{eqnarray}
K_{(b1)}(p,\kappa;E) \!&=&\! \frac{1}{2p \kappa}\ln\bigg[\frac{\frac{M_\Lambda}{2\mu_{\Lambda T}}p^{2}+\kappa^2+p \kappa-M_\Lambda E}{\frac{M_\Lambda}{2\mu_{\Lambda T}}p^{2}+\kappa^2-p \kappa-M_\Lambda E}\bigg]\,,
\nonumber\\
\nonumber\\
K_{(b2)}(p,\kappa;E) \!&=&\! \frac{1}{2p \kappa}\ln\bigg[\frac{p^{2}+\frac{M_\Lambda}{2\mu_{\Lambda T}}\kappa^2+p \kappa-M_\Lambda E }{p^{2}+\frac{M_\Lambda}{2\mu_{\Lambda T}}\kappa^2-p \kappa-M_\Lambda E}\bigg]\,,\,\,\,\,\quad\,
\end{eqnarray}
respectively, where the generic momentum $\kappa\, =k\,(q)$ denotes the on-shell (loop) momenta. The inclusion 
of the regulator-dependent ($\Lambda_c$-dependent) three-body contact couplings $g^{(A,B)}_3(\Lambda_c)$ modifies the 
one-particle exchange interaction kernels, $K_{(a)}$ and $K_{(b2)}$, in the respective elastic channels as: 
\begin{eqnarray}
\mathcal{K}^{A,B}_{(a)}(p,\kappa,\Lambda_c;E) &=& \left[K_{(a)}(p,\kappa;E)
-\frac{g^{(A,B)}_3(\Lambda^2_c)}{\Lambda^2_c}\right]\,,
\nonumber\\
\nonumber\\
\mathcal{K}^{A,B}_{(b2)}(p,\kappa,\Lambda_c;E) &=& \left[K_{(b2)}(p,\kappa;E)
-\frac{g^{(A,B)}_3(\Lambda^2_c)}{\Lambda^2_c}\right].\,\quad
\label{eq:g3_modified_K}
\end{eqnarray}
Here we point out that in this work we used a minimal prescription of introducing the scale dependent three-body 
couplings only in the elastic channels. In general, the most systematic method of renormalization is to include 
them in all the inelastic channels as well, e.g., as done in Refs.~\cite{Hammer:2001ng,Hildenbrand:2019sgp}. 
In the present case we find that the latter method leads to certain uncontrollable numerical instabilities in 
determining the limit cycle behaviors of $g^{(A,B)}_3(\Lambda_c)$. This is perhaps due to the simultaneous 
admixture of the negative $(\Lambda\Lambda)_s$ and positive $(\Lambda T)_{s,t}$ two-body scattering lengths 
associated with the virtual and real bound state dimers, respectively. Hence, we took recourse to the former 
simplistic prescription. Either way, since these unknown scale dependent three-body couplings are needed to be 
fixed phenomenologically during evaluations of the integral equations, they are expected to get accordingly 
renormalized in the different coupled channels. Thereby, the essential qualitative features of our investigations 
of the three-body bound states (e.g., the quasiperiodicity of the RG limit cycle) are by and large expected to 
remain unaffected. This issue is elucidated later in our results presented in the forthcoming section.

%%%%%%%%%%%%%%%%%%%%%%%%%%%%%%%
\subsection{Three-body scattering lengths}
%%%%%%%%%%%%%%%%%%%%%%%%%%%%%%%
The coupled integral equations displayed in the previous subsection must be renormalized and then solved 
numerically to yield predictions for the $\Lambda\Lambda T$ three-body scattering amplitudes. For a given 
on-shell relative momentum $k=|{\bf k}|$ and three-body center-of-mass kinetic energy $E$, the kinematical 
scattering domain lies between the particle-dimer breakup thresholds ${\mathcal E}^{thr}_{2(s,t)}$ and 
the three-particle breakup threshold, i.e., ${\mathcal E}^{thr}_{2(s,t)}<E<0$. In contrast with the 
kinematical domain of three-body bound states ($E<{\mathcal E}^{thr}_{2(s,t)}$ with imaginary $k$) free of 
singularities, the integral equations in the scattering domain develop singularities associated with poles
of the $(\Lambda T)_{s,t}$-dimer propagators ${\mathcal D}_{0,1}(q,E)$ for certain values of the loop 
momenta $q$. For the type-A integral equations the only poles are those that arise from the 
${\mathcal D}_{0}(q,E)$ propagator insertions at $q=k$. While for the type-B integral equations poles arise 
due to the insertions of both $(\Lambda T)_{s,t}$-dimer propagators, namely, ${\mathcal D}_{1}(q,E)$ has a 
pole at $q=k$ and ${\mathcal D}_{0}(q,E)$ has a pole at 
$q= \sqrt{k^2+ (\gamma_0^2-\gamma_1^2)(\mu_{\Lambda(\Lambda T)}/\mu_{\Lambda T})}$. To avoid these poles, a 
{\it principal value} prescription must be applied in the appropriate loop integrals to extract the 
three-body scattering amplitudes. Furthermore, it is numerically advantageous to express the otherwise 
complex-valued integral equations below the three-particle breakup threshold in terms of the real-valued 
renormalized $K$-matrix elements ${\mathbb K}^{(A,B)}_{a,b,c}(p,k;E)$ for the respective choice of the elastic 
processes, viz. $u_{(0,1)}\Lambda\to u_{(0,1)}\Lambda$. To this end we display the principal value 
prescription modified renormalized $K$-matrix integral equations: 
\begin{widetext}
%%%%%%%%%%%%%%%%%%%%%%%
\begin{eqnarray}
{\mathbb K}^{(A)}_a(p,k;E) 
&=& -\frac{M_T}{4\mu_{\Lambda T}}\mathcal{M}^{A(0)}_{(a)}(p,k;E)
-\frac{M_T}{2\pi\mu_{\Lambda T}}\,{\mathcal P}\!\!\int_0^{\Lambda_c} dq\,{\mathcal M}^{A(0)}_{(a)}(p,q,\Lambda_c;E)\,\frac{q^2}{q^2-k^2}\,{{\mathbb K}^{(A)}_a(q,k;E)}
\nonumber\\
\nonumber\\
&&+\,\frac{\sqrt{3}M_T}{2\pi\mu_{\Lambda T}}\frac{y_0}{y_1}\int_0^{\Lambda_c} dq\,{\mathcal M}^{A(0)}_{(a)}(p,q,\Lambda_c;E)\,\frac{q^2}{q^2-k^2+\frac{\mu_{\Lambda(\Lambda T)}}{\mu_{\Lambda T}}(\gamma^2_0-\gamma^2_1)}\,{{\mathbb K}^{(A)}_b(q,k;E)}
\nonumber\\
\nonumber\\
&&-\,\frac{\sqrt{2}}{\pi}\,\frac{y_0}{y_s}\,{\mathcal P}\!\!\int_0^{\Lambda_c} dq\,{\mathcal M}^{A(0)}_{(b2)}(p,q,\Lambda_c;E)\,\frac{q^2}{q^2-k^2}\,{{\mathbb K}^{(A)}_c(q,k;E)}\,,
\nonumber
\end{eqnarray}
%%%%%%%%%%%%%%%%%%%%%%%
\begin{eqnarray}
{\mathbb K}^{(A)}_b(p,k;E) 
&=& \frac{\sqrt{3}M_T}{4\mu_{\Lambda T}}\frac{y_1}{y_0}\,M^{(1)}_{(a)}(p,k;E)
+\frac{\sqrt{3}M_T}{2\pi\mu_{\Lambda T}}\frac{y_1}{y_0}\,{\mathcal P}\!\!\int_0^{\Lambda_c} dq\,M^{(1)}_{(a)}(p,q;E)\,\frac{q^2}{q^2-k^2}\,{{\mathbb K}^{(A)}_a(q,k;E)}
\nonumber\\
\nonumber\\
&&+\,\frac{M_T}{2\pi\mu_{\Lambda T}}\int_0^{\Lambda_c} dq\, M^{(1)}_{(a)}(p,q;E)\,\frac{q^2}{q^2-k^2+\frac{\mu_{\Lambda(\Lambda T)}}{\mu_{\Lambda T}}(\gamma^2_0-\gamma^2_1)}\,{{\mathbb K}^{(A)}_b(q,k;E)}
\nonumber\\
\nonumber\\
&&-\,\frac{\sqrt{6}}{\pi}\,\frac{y_1}{y_s}\,{\mathcal P}\!\!\int_0^{\Lambda_c} dq\, M^{(1)}_{(b2)}(p,q;E)\,\frac{q^2}{q^2-k^2}\,{{\mathbb K}^{(A)}_c(q,k;E)}\,,
\nonumber\\
\nonumber\\
\nonumber\\
%%%%%%%%%%%%%%%%%%%%%%%
{\mathbb K}^{(A)}_c(p,k;E) 
&=& \frac{M_\Lambda}{2\sqrt{2}\mu_{\Lambda T}}\frac{y_s}{y_0}\,M_{(b1)}(p,k;E)
+\frac{M_\Lambda}{\sqrt{2}\pi\mu_{\Lambda T}}\frac{y_s}{y_0}\,{\mathcal P}\!\!\int_0^{\Lambda_c} dq\, M_{(b1)}(p,q;E)\,\frac{q^2}{q^2-k^2}\,{{\mathbb K}^{(A)}_a(q,k;E)}
\nonumber\\
\nonumber\\
&&\,+\,\sqrt{\frac{3}{2}}\frac{M_\Lambda}{\pi\mu_{\Lambda T}}\frac{y_s}{y_1}\,\int_0^{\Lambda_c} dq\, M_{(b1)}(p,q;E)\,\frac{q^2}{q^2-k^2+\frac{\mu_{\Lambda(\Lambda T)}}{\mu_{\Lambda T}}(\gamma^2_0-\gamma^2_1)}\,{{\mathbb K}^{(A)}_b(q,k;E)}\,,
\nonumber\\
\label{eq:type-KA}
\end{eqnarray}
%%%%%%%%%%%%%%%%%%%%%%%
for the type-A elastic channel with $E\equiv E_A={\mathcal E}^{thr}_{2(s)}+k^2/(2\mu_{\Lambda(\Lambda T)})$, and \\
%%%%%%%%%%%%%%%%%%%%%%%
\begin{eqnarray}
{\mathbb K}^{(B)}_a(p,k;E) 
&=& \frac{M_T}{4\mu_{\Lambda T}}{\mathcal M}^{B(1)}_{(a)}(p,k;E) 
+\frac{M_T}{2\pi\mu_{\Lambda T}}\,{\mathcal P}\!\!\int_0^{\Lambda_c} dq\, {\mathcal M}^{B(1)}_{(a)}(p,q,\Lambda_c;E)\,\frac{q^2}{q^2-k^2}\,{{\mathbb K}^{(B)}_a(q,k;E)}
\nonumber\\
\nonumber\\
&&+\,\frac{\sqrt{3}M_T}{2\pi\mu_{\Lambda T}}\frac{y_1}{y_0}\,{\mathcal P}\!\!\int_0^{\Lambda_c} dq\,{\mathcal M}^{B(1)}_{(a)}(p,q,\Lambda_c;E)\,\frac{q^2}{q^2-k^2-\frac{\mu_{\Lambda(\Lambda T)}}{\mu_{\Lambda T}}(\gamma^2_0-\gamma^2_1)}\,{{\mathbb K}^{(B)}_b(q,k;E)}
\nonumber\\
\nonumber\\
&&-\,\frac{\sqrt{6}}{\pi}\,\frac{y_1}{y_s}\,{\mathcal P}\!\!\int_0^{\Lambda_c} dq\,{\mathcal M}^{B(1)}_{(b2)}(p,q,\Lambda_c;E)\,\frac{q^2}{q^2-k^2}\,{{\mathbb K}^{(B)}_c(q,k;E)}\,\,,
\nonumber\\
\nonumber\\
\nonumber\\
%%%%%%%%%%%%%%%%%%%%%%%
{\mathbb K}^{(B)}_b(p,k;E) 
&=&  \frac{\sqrt{3}M_T}{4\mu_{\Lambda T}}\frac{y_0}{y_1}\,M^{(0)}_{(a)}(p,k;E)
+\frac{\sqrt{3}M_T}{2\pi\mu_{\Lambda T}}\frac{y_0}{y_1}\,{\mathcal P}\!\!\int_0^{\Lambda_c} dq\, M^{(0)}_{(a)}(p,q;E)\,\frac{q^2}{q^2-k^2}\,{{\mathbb K}^{(B)}_a(q,k;E)}
\nonumber\\
\nonumber\\
&&-\,\frac{M_T}{2\pi\mu_{\Lambda T}}\,{\mathcal P}\!\!\int_0^{\Lambda_c} dq\, M^{(0)}_{(a)}(p,q;E)\,\frac{q^2}{q^2-k^2-\frac{\mu_{\Lambda(\Lambda T)}}{\mu_{\Lambda T}}(\gamma^2_0-\gamma^2_1)}\,{{\mathbb K}^{(B)}_b(q,k;E)}
\nonumber\\
\nonumber\\
&&-\,\frac{\sqrt{2}}{\pi}\,\frac{y_0}{y_s}\,{\mathcal P}\!\!\int_0^{\Lambda_c} dq\, M^{(0)}_{(b2)}(p,q;E)\,\frac{q^2}{q^2-k^2}\,{{\mathbb K}^{(B)}_c(q,k;E)}\,\,,
\nonumber\\
\nonumber\\
\nonumber\\
%%%%%%%%%%%%%%%%%%%%%%%
{\mathbb K}^{(B)}_c(p,k;E) 
&=& \frac{\sqrt{3}M_\Lambda}{2\sqrt{2}\mu_{\Lambda T}}\frac{y_s}{y_1}\,M_{(b1)}(p,k;E)
+\,\sqrt{\frac{3}{2}}\,\frac{M_\Lambda}{\pi\mu_{\Lambda T}}\frac{y_s}{y_1}\,{\mathcal P}\!\!\int_0^{\Lambda_c} dq\, M_{(b1)}(p,q;E)\,\frac{q^2}{q^2-k^2}\,{{\mathbb K}^{(B)}_a(q,k;E)}
\nonumber\\
\nonumber\\
&&+\,\frac{M_\Lambda}{\sqrt{2}\pi\mu_{\Lambda T}}\frac{y_s}{y_0}\,{\mathcal P}\!\!\int_0^{\Lambda_c} dq\, M_{(b1)}(p,q;E)\,\frac{q^2}{q^2-k^2-\frac{\mu_{\Lambda(\Lambda T)}}{\mu_{\Lambda T}}(\gamma^2_0-\gamma^2_1)}\,{{\mathbb K}^{(B)}_b(q,k;E)}\,,
\nonumber\\
\label{eq:type-KB}
\end{eqnarray}
%%%%%%%%%%%%%%%%%%%%%%%
\end{widetext}
for the type-B elastic channel with $E\equiv E_B={\mathcal E}^{thr}_{2(t)}+k^2/(2\mu_{\Lambda(\Lambda T)})$. 
The symbol ``$\mathcal P$'' stands for a principal value integral which involves rewriting the 
complex-valued dimer propagators with $i\eta$ prescription in terms of real-valued propagators, namely,
\begin{equation*}
\frac{1}{q^2-k^2-i\eta}={\mathcal P}\frac{1}{q^2-k^2}+i\pi\delta(q^2-k^2)\,,
\end{equation*}
and 
\begin{eqnarray}
\frac{1}{q^2-k^2-\frac{\mu_{\Lambda(\Lambda T)}}{\mu_{\Lambda T}}(\gamma^2_0-\gamma^2_1) - i\eta} &&
\nonumber\\
&&\hspace{-4cm}={\mathcal P}\frac{1}{q^2-k^2-\frac{\mu_{\Lambda(\Lambda T)}}{\mu_{\Lambda T}}(\gamma^2_0-\gamma^2_1)}
\nonumber\\
&&\hspace{-3.5cm}+\,i\pi\delta\left(q^2-k^2-\frac{\mu_{\Lambda(\Lambda T)}}{\mu_{\Lambda T}}(\gamma^2_0-\gamma^2_1)\right)\,.
\nonumber
\end{eqnarray}
The $S$-wave projected $\Lambda$ and $T$-exchange interactions kernels in this case are rewritten as:
\begin{eqnarray}
M^{(0,1)}_{(a)}(p,\kappa;E)&=&\left(\frac{\mu_{\Lambda(\Lambda T)}}{\mu_{\Lambda T}}\right)K_{(a)}(p,\kappa;E)
\nonumber\\
&&\times\,\left(\gamma_{0,1}+\sqrt{p^2\frac{\mu_{\Lambda T}}{\mu_{\Lambda (\Lambda T)}}-2\mu_{\Lambda T}E}\right)\,,
\nonumber\\
M_{(b1)}(p,\kappa;E)&=&K_{(b1)}(p,\kappa;E)
\nonumber\\
&&\times\,\left(\frac{p^2-k^2}{\frac{1}{a_{\Lambda\Lambda}}-\sqrt{p^2\frac{M_\Lambda}{2\mu_{T(\Lambda\Lambda )}}-M_\Lambda E}}\right)\,,
\nonumber\\
M^{(0,1)}_{(b2)}(p,\kappa;E)&=&\left(\frac{\mu_{\Lambda(\Lambda T)}}{\mu_{\Lambda T}}\right)K_{(b2)}(p,\kappa;E)
\nonumber\\
&&\times\,\left(\gamma_{0,1}+\sqrt{p^2\frac{\mu_{\Lambda T}}{\mu_{\Lambda (\Lambda T)}}-2\mu_{\Lambda T}E}\right)\,,
\nonumber\\
\end{eqnarray}
and the corresponding three-body force modified $\Lambda_c$ dependent kernels needed are:
\begin{eqnarray}
{\mathcal M}^{A,B(0,1)}_{(a)}(p,\kappa,\Lambda_c;E)&=&\left(\frac{\mu_{\Lambda(\Lambda T)}}{\mu_{\Lambda T}}\right){\mathcal K}^{A,B}_{(a)}(p,\kappa,\Lambda_c;E)
\nonumber\\
&\times&\left(\gamma_{0,1}+\sqrt{p^2\frac{\mu_{\Lambda T}}{\mu_{\Lambda (\Lambda T)}}-2\mu_{\Lambda T}E}\right)\,,
\nonumber\\
{\mathcal M}^{A,B(0,1)}_{(b2)}(p,\kappa,\Lambda_c;E)&=&\left(\frac{\mu_{\Lambda(\Lambda T)}}{\mu_{\Lambda T}}\right){\mathcal K}^{A,B}_{(b2)}(p,\kappa,\Lambda_c;E)
\nonumber\\
&\times&\left(\gamma_{0,1}+\sqrt{p^2\frac{\mu_{\Lambda T}}{\mu_{\Lambda (\Lambda T)}}-2\mu_{\Lambda T}E}\right)\,,
\nonumber\\
\end{eqnarray}
where $\kappa=k\,(q)$ is the on-shell (loop) momentum. In the above integral equations, the unrenormalized 
complex-valued amplitudes $T_a^{(A,B)}(p,k;E)$ are related to the renormalized real-valued $K$-matrix elements 
${\mathbb K}^{(A,B)}_{a,b,c}(p,k;E)$ by the following relations:
\begin{equation*}
\frac{{\mathbb K}^{(A)}_{a}(p,k;E)}{k^2-p^2}=\left(\frac{\mu_{\Lambda T}}{4\pi \gamma_0}\right)\frac{\sqrt{Z_{0}} \,T_{a}^{(A)}(p,k;E) \sqrt{Z_{0}}}{\gamma_0-\sqrt{q^2\frac{\mu_{\Lambda T}}{\mu_{\Lambda(\Lambda T)}}-2\mu_{\Lambda T} E}}\,,
\end{equation*}

\vspace{-0.6cm}

\begin{eqnarray}
\frac{{\mathbb K}^{(A)}_{b}(p,k;E)}{k^2-p^2-\frac{\mu_{\Lambda(\Lambda T)}}{\mu_{\Lambda T}}(\gamma^2_0-\gamma^2_1)}&&
\nonumber\\
&&\hspace{-2cm}=\,\left(\frac{\mu_{\Lambda T}}{4\pi \gamma_0}\right)\frac{\sqrt{Z_{0}} \,T_{b}^{(A)}(p,k;E) \sqrt{Z_{0}}}{\gamma_1-\sqrt{q^2\frac{\mu_{\Lambda T}}{\mu_{\Lambda(\Lambda T)}}-2\mu_{\Lambda T} E}}\,,
\nonumber
\end{eqnarray}

\vspace{-0.6cm}

\begin{eqnarray}
\frac{{\mathbb K}^{(A)}_{c}(p,k;E)}{k^2-p^2}&=&\left(\frac{\mu_{\Lambda T}}{4\pi \gamma_0}\right)\frac{\sqrt{Z_{0}} \,T_{c}^{(A)}(p,k;E) \sqrt{Z_{0}}}{\frac{1}{a_{\Lambda\Lambda}}-\sqrt{q^2\frac{M_{\Lambda}}{2\mu_{T(\Lambda\Lambda)}}-M_{\Lambda} E}}\,,
\nonumber\\
\label{eq:renorm_aR-SetA}
\end{eqnarray}
for the type-A amplitudes, and
\begin{equation*}
\frac{{\mathbb K}^{(B)}_{a}(p,k;E)}{k^2-p^2}=\left(\frac{\mu_{\Lambda T}}{4\pi \gamma_1}\right)\frac{\sqrt{Z_{1}} \,T_{a}^{(B)}(p,k;E) \sqrt{Z_{1}}}{\gamma_1-\sqrt{q^2\frac{\mu_{\Lambda T}}{\mu_{\Lambda(\Lambda T)}}-2\mu_{\Lambda T} E}}\,,
\end{equation*}
\begin{eqnarray}
\frac{{\mathbb K}^{(B)}_{b}(p,k;E)}{k^2-p^2+\frac{\mu_{\Lambda(\Lambda T)}}{\mu_{\Lambda T}}(\gamma^2_0-\gamma^2_1)}&&
\nonumber\\
&&\hspace{-1.7cm}=\,\left(\frac{\mu_{\Lambda T}}{4\pi \gamma_1}\right)\frac{\sqrt{Z_{1}} \,T_{b}^{(B)}(p,k;E) \sqrt{Z_{1}}}{\gamma_0-\sqrt{q^2\frac{\mu_{\Lambda T}}{\mu_{\Lambda(\Lambda T)}}-2\mu_{\Lambda T} E}}\,,
\nonumber\\
\frac{{\mathbb K}^{(B)}_{c}(p,k;E)}{k^2-p^2}=\left(\frac{\mu_{\Lambda T}}{4\pi \gamma_1}\right)&&\frac{\sqrt{Z_{1}} \,T_{c}^{(B)}(p,k;E) \sqrt{Z_{1}}}{\frac{1}{a_{\Lambda\Lambda}}-\sqrt{q^2\frac{M_{\Lambda}}{2\mu_{T(\Lambda\Lambda)}}-M_{\Lambda} E}}\,,
\nonumber\\
\label{eq:renorm_aR-SetB}
\end{eqnarray}
for the type-B amplitudes, where $Z_{0,1}$ are the $u_{0,1}$-dimer field wave function renormalization 
constants, defined as the residues of the renormalized dressed dimer propagators 
$\Delta_{0,1}(k_0,{\bf k}) $ [cf. Eq.~\eqref{dimer_props} in the appendix]:
\begin{eqnarray}
Z_{0}^{-1} &=& \frac{d[\Delta^{-1}_{0}(k_0,{\bf 0})]}{dk_0}\bigg|_{k_0=-\mathcal{B}_{\Lambda}[0^+]} 
= \frac{\mu_{\Lambda T}^2y_0^2}{2\pi \gamma_0}\,,
\nonumber\\
Z_{1}^{-1} &=& \frac{d[\Delta^{-1}_{1}(k_0,{\bf 0})]}{dk_0}\bigg|_{k_0=-\mathcal{B}_{\Lambda}[1^+]} 
= \frac{\mu_{\Lambda T}^2y_1^2}{2\pi \gamma_1}\,.
\end{eqnarray}
Finally, the $J=1/2$ $S$-wave $\Lambda\Lambda T$ scattering lengths corresponding to the constituent 
spin-singlet and spin-triplet $\Lambda T$ subsystems are obtained by numerically solving the above 
$K$-matrix equations for the renormalized on-shell elastic-scattering amplitudes 
${\mathbb K}^{(A,B)}_a(k,k)$, and then taking the threshold limit according to the definition 
\begin{equation}
a_{3(s,t)} =-\lim_{k\to 0} {\mathbb K}^{(A,B)}_a(k,k)\,.
\label{a3st}
\end{equation}
It is notable that neither of the two three-body scattering lengths $a_{3(s,t)}$ can be considered as 
physical observables. On the other hand, albeit practical difficulties, it may not be on the whole 
impossible to extract the effective three-body scattering length $a_{\Lambda\Lambda T}$ 
at low-energies from the $(2J+1)$-{\it spin averaged} $S$-wave elastic cross section 
$\sigma^{el}_{\Lambda\Lambda T}$ by using the relation
\begin{equation}
\label{eq:aLL}
a_{\Lambda\Lambda T}=\sqrt{\frac{1}{4}a_{3(s)}^2+\frac{3}{4}a_{3(t)}^2}\,\,, 
\end{equation}
 vis-a-vis, the prescription: 
\begin{eqnarray}
\sigma^{el}_{\Lambda\Lambda T}&=&\frac{1}{4}\sigma_{3(s)}(\text{type-A})+\frac{3}{4}\sigma_{3(t)}(\text{type-B})\,;
\nonumber\\
a_{3(s,t)}&=&\lim_{k\to 0}\sqrt{\frac{1}{4\pi}\sigma_{3(s,t)}(\text{type-A,B})}\,,
\nonumber\\
a_{\Lambda\Lambda T}&=&\lim_{k\to 0}\sqrt{\frac{1}{4\pi}\sigma^{el}_{\Lambda\Lambda T}}\,.
\end{eqnarray}
Thus, our EFT framework provides a viable prescription to determine the three-body scattering lengths 
via numerical solutions to the renormalized $K$-matrix integral equations. Having said that it 
must be borne in mind that as yet there exists no experimental facility capable of extracting these 
scattering lengths by measuring the above elastic cross sections. The unstable nature of the 
$\Lambda$-hyperon poses immense technical challenges to be used either as targets or projectiles in 
scattering experiments. Nevertheless, the purpose of the present exercise is to demonstrate the kind 
of prototypical analysis that may be necessary whenever such information becomes available from future 
experimental investigations.   

%%%%%%%%%%%%%%%%%%%%%%%%%%%%%%%
\subsection{Asymptotic bound state analysis}
%%%%%%%%%%%%%%%%%%%%%%%%%%%%%%%
In the investigation of three-body bound state characteristic in the $\Lambda\Lambda T$ cluster systems, 
the emergence of RG limit-cycle behavior could be easily checked by studying the UV limit of the coupled 
integral equations where the off-shell or loop momenta is asymptotically large, i.e., 
$q,p\sim\Lambda_c\to\infty$, while the on-shell energy and relative momenta is small, i.e., 
$E,k\sim \gamma_{0,1}\sim 1/a_{\Lambda\Lambda}\ll p, q$. In this limit the inhomogeneous parts as well
as the $\Lambda^{-2}_c$ suppressed three-body contributions to the integral equations drop out. After 
suitable redefinitions of the half-off-shell amplitudes, they may be shown to scale for generic 
off-shell asymptotic momenta $\kappa$ as $T^{(A,B)}_{a,b,c}(\kappa\to\infty )\sim \kappa^{s-1}$. Finally
through a sequence of {\it Mellin transformations}, both sets of integral equations reduce to same 
transcendental form: 
\begin{eqnarray}
1&=&\bigg(\frac{M_T}{2\pi\mu_{\Lambda T}C_1}\bigg)
\bigg[\frac{2\pi}{s}\frac{\sin\left[s\sin^{-1}(a/2)\right]}{\cos[\pi s/2]}\bigg]
\nonumber\\
&&+\,\bigg(\frac{M_\Lambda}{\pi^2\mu_{\Lambda T}C_1C_2}\bigg)
\bigg[ \frac{2\pi }{s}\frac{\sin\left[s \cot^{-1}\sqrt{4b-1}\,\right]}{\cos[\pi s/2]}\bigg]^2, \,\quad
\label{eq:asymptotic}
\end{eqnarray}
where 
\begin{eqnarray}
a=\frac{2\mu_{\Lambda T}}{M_T}\,\,&,&\,\, b=\frac{M_\Lambda}{2\mu_{\Lambda T}}\,,
\nonumber\\ 
C_1=\sqrt{\frac{\mu_{\Lambda T}}{\mu_{\Lambda(\Lambda T)}}}\,\,&,&\,\, 
C_2=\sqrt{\frac{M_\Lambda}{2\mu_{T(\Lambda\Lambda )}}}\,.
\nonumber 
\end{eqnarray}
Solving for the exponent $s$ in above equation yields the following imaginary values:
\begin{equation}
s =\pm\, is_{0}^{\infty}
\begin{cases}
 s_{0}^{\infty}=1.03517... & \text{for }\, {}_{\Lambda\Lambda}^{\,\,\,\,5}{\rm H} \\ 
 s_{0}^{\infty}=1.03516... & \text{for }\, {}_{\Lambda\Lambda}^{\,\,\,\,5}{\rm He}\,. 
\end{cases}
\label{eq:s0_asy}
\end{equation}
The small numerical difference between the values of the asymptotic limit cycle parameter $s^{\infty}_0$ 
reflects their universal character with reasonably good isospin symmetry in the three-body sector. The 
imaginary solutions can be formally attributed to the existence of Efimov states in the {\it unitary limit} 
of the two mirror $\Lambda\Lambda T$ clusters and parametrize the onset of discrete scaling invariance. A 
detailed exposition of this kind of asymptotic analysis leading to the Efimov effect is found in 
Ref.~\cite{Braaten:2004rn}. In the next section we present a qualitative assay of our numerical results for
the nonasymptotic solutions to the integral equations and their possible implications in the low-energy 
domain. 

%%%%%%%%%%%%%%%%%%%%%%%%%%%%%TABLE1%%%%%%%%%%%%%%%%%%%%%%%%%%%%%%
\begin{table}[tbp]
\begin{tabular}{ |c||c|c|c| }
\hline\hline
Particle           & Symbol       & Mass (MeV) &  Binding energy (MeV) \\
\hline\hline
$\Lambda$-hyperon  & $\Lambda$    & 1115.683   &    -                  \\
Triton ${}^3$H     & $t$          & 2808.921   &    8.48               \\
Helion ${}^3$He    & $h$          & 2808.391   &    7.72               \\
\hline\hline 
\end{tabular}
\caption{Particle data used in our calculations~\cite{Mohr:2015ccw}.}
\label{table:T1}
\end{table}
%%%%%%%%%%%%%%%%%%%%%%%%%%%%%%%%%%%%%%%%%%%%%%%%%%%%%%%%%%%%%%%%%

%%%%%%%%%%%%%%%%%%%%%%%%%%%%%%%%%%%%%%%%%%%%%%%%%%%%%%%%%%%%%%%%%%%%%%%%
\section{RESULTS AND DISCUSSION}\label{sec:results}
%%%%%%%%%%%%%%%%%%%%%%%%%%%%%%%%%%%%%%%%%%%%%%%%%%%%%%%%%%%%%%%%%%%%%%%%

%%%%%%%%%%%%%%%%%%%%%%%%%%%%TABLE2%%%%%%%%%%%%%%%%%%%%%%%%%%%%%%%
\begin{table*}[t]
\begin{tabular}{ |c||c|c||c|c|c| }
\hline\hline
$\Lambda\Lambda$-Hypernuclear              & $S$-wave $\Lambda\Lambda$                                    & $\Lambda\Lambda$-Separation  & Incremental binding                                  & Critical cutoff                    & cutoff                           \\ 
mirror (a\,, b)                            &  Scattering length                                         &  energy $B_{\Lambda\Lambda}$ & energy $\Delta B_{\Lambda\Lambda}$ (MeV)   & $\Lambda^{(n=0)}_{\rm crit}$ (MeV)  & $\Lambda^{(n=0)}_{\rm pot}$ (MeV) \\
Sets                                       &  $a_{\Lambda\Lambda}$ (fm)                                 & (MeV) \cite{Nemura:2004xb}   & reevaluated (this work) & (with $g^{(A,B)}_3=0$ )             & (with $g^{(A,B)}_3=0$)            \\
\hline\hline
Ia (${}_{\Lambda\Lambda}^{\,\,\,\,5}$H)    & \textcolor{red}{\bf -0.91} (mND$_S$)~\cite{Nagels:1978sc}  & \textcolor{red}{\bf 3.750}   & 1.071                                                &  235.028                            & 437.654                           \\
Ib (${}_{\Lambda\Lambda}^{\,\,\,\,5}$He)   & \textcolor{red}{\bf -0.91} (mND$_S$)~\cite{Nagels:1978sc}  & \textcolor{red}{\bf 3.660}   & 0.989                                                &  269.621                            &  429.833                          \\  
\hline
IIa (${}_{\Lambda\Lambda}^{\,\,\,\,5}$H)   & -1.37 (ND$_S$)~\cite{Nagels:1978sc}                        &  4.050                       & 1.381                                                &  205.448                            & 403.285                           \\
IIb (${}_{\Lambda\Lambda}^{\,\,\,\,5}$He)  & -1.37 (ND$_S$)~\cite{Nagels:1978sc}                        & 3.960                        & 1.289                                                & 234.522                             & 396.332                           \\
\hline\hline  
\end{tabular}
\caption{Two sets of predictions for the three-body binding or double-$\Lambda$-separation energy $B_{\Lambda\Lambda}$ for
         the (${}_{\Lambda\Lambda}^{\,\,\,\,5}{\rm H}\,,\,{}_{\Lambda\Lambda}^{\,\,\,\,5}{\rm He}$) mirrors using the coupled-channel potential model SVM analysis of Nemura {\it et al.}~\cite{Nemura:2004xb}. The corresponding double-$\Lambda$ 
         scattering lengths used are two representative values based on the old Nijmegen hard-core potential
         models~\cite{Nagels:1978sc} (names in parentheses) consistent with the currently accepted range, 
         $-1.92~{\rm fm} \lesssim a_{\Lambda\Lambda}\lesssim -0.5~{\rm fm}$~\cite{Morita:2014kza,Ohnishi:2015cnu,Ohnishi:2016elb}, 
         as constrained by the recent RHIC data~\cite{Adamczyk:2014vca}. The values of the incremental binding energies 
         $\Delta B_{\Lambda\Lambda}$ are obtained utilizing the recent experimental input for the $\Lambda$-separation energies 
         of the ground (singlet) and first (triplet) excited states of the 
         (${}_{\Lambda}^{4}{\rm H}\,,\,{}_{\Lambda}^{4}{\rm He}$) 
         mirrors~\cite{Esser:2015trs,Schulz:2016kdc,Yamamoto:2015avw,Koike:2019rrs}. Furthermore, with the three-body contact 
         interactions excluded from our integral equations, the critical cutoffs, $\Lambda_c=\Lambda^{(n=0)}_{\rm crit}$ (see
         text), associated with the ground ($n=0$) state Efimov-like trimers for each mirror double-$\Lambda$-hypernuclei, are 
         also displayed. The rightmost column shows our adjusted cutoff values, $\Lambda_c=\Lambda^{(n=0)}_{\rm pot}$, which 
         reproduce the above values of $B_{\Lambda\Lambda}$ as ground state eigenenergies. The paired 
         ($B_{\Lambda\Lambda}\,,\, a_{\Lambda\Lambda}$) data points for cases Ia and Ib (shown in bold) are 
         used to normalize our solutions. }
\label{table:T2}    
\end{table*}
%%%%%%%%%%%%%%%%%%%%%%%%%%%%%%%%%%%%%%%%%%%%%%%%%%%%%%%%%%%%%%%%%
%--figure---------------------------------
\begin{figure*}[tbp]
    \centering
    \includegraphics[width=\columnwidth]{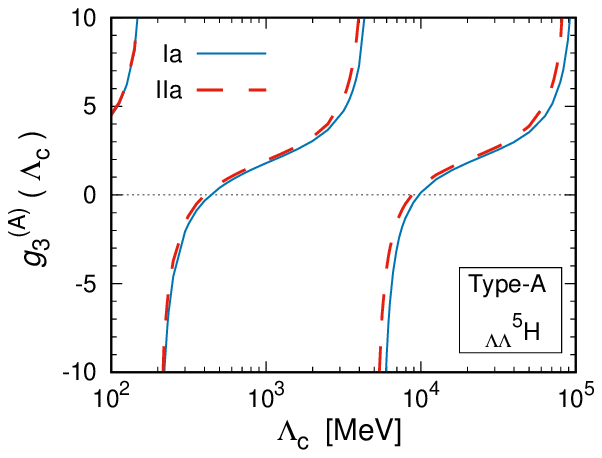}~~\includegraphics[width=\columnwidth]{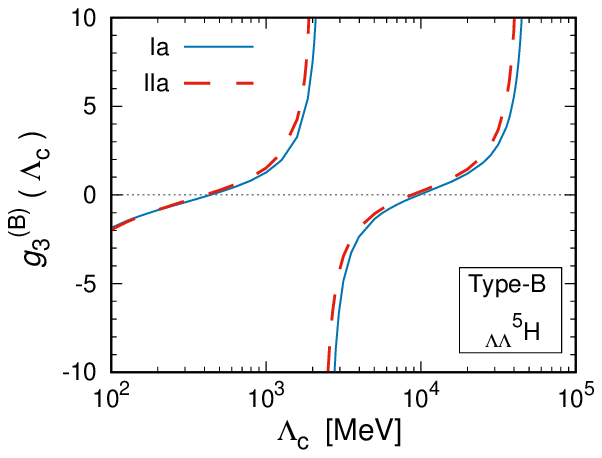}
    \caption{\label{fig:fig3} The nonasymptotic RG limit cycle behaviors of the three-body couplings 
             $g^{(A,B)}_3(\Lambda_c)$ for the $\Lambda\Lambda t$ system. Two representative choices 
             for the $S$-wave double-$\Lambda$ scattering lengths are considered, namely, 
             $a_{\Lambda\Lambda}=-0.91~{\rm fm\,(Ia)}$ and $-1.37~{\rm fm\,(IIa)}$, based on the 
             Nijmegen hard-core potential models, mND$_S$ and ND$_S$, respectively~\cite{Nagels:1978sc}, 
             and compatible with the range of values constrained by the recent phenomenological 
             analyses~\cite{Morita:2014kza,Ohnishi:2015cnu,Ohnishi:2016elb} of RHIC 
             data~\cite{Adamczyk:2014vca}. The corresponding three-body binding or 
             double-$\Lambda$-separation energies $B_{\Lambda\Lambda}$ (cf. Table~\ref{table:T2}) used as
             input to our integral equations, are the predictions of the {\it ab initio } potential model 
             analysis of Ref.~\cite{Nemura:2004xb}. The corresponding results for the $\Lambda\Lambda h$ 
             system being almost identical are not displayed for brevity.}
\end{figure*}
%--figure---------------------------------
 
For our numerical evaluations, we use the masses of the particles as displayed in Table.~\ref{table:T1}. 
As a comparison with our already obtained asymptotic limit cycle parameter $s_{0}^{\infty}$ for each mirror 
hypernuclei, the analogous nonasymptotic parameter $s_{0}$ may be obtained by studying the RG behavior 
of the three-body couplings $g^{(A,B)}_3(\Lambda_c)$ for nonasymptotic kinematics. The $s_0$ parameter 
is, however, nonuniversal in character and sensitive to the cutoff variations. Nevertheless, it may be 
shown that as $\Lambda_c\to \infty$, $s_{0}\to s_{0}^{\infty}$~\cite{Raha:2017ahu}. We note that currently 
there is no empirical three-body information available to constraint $g^{(A,B)}_3$. Thus, we adopt a 
strategy similar to the earlier pursued works~\cite{Ando:2013kba,Ando:2015uda,Raha:2017ahu}. We assume that 
${}_{\Lambda\Lambda}^{\,\,\,\,5}$H and ${}_{\Lambda\Lambda}^{\,\,\,\,5}$He already form Efimov-like bound 
cluster states and thereby investigate the RG of $g^{(A,B)}_3$ by choosing two sets of values of the 
three-body binding or double-$\Lambda$-separation energies\footnote{The double-$\Lambda$-separation energy 
$B_{\Lambda\Lambda}$, as commonly referred to in the context of potential model analyses, is interpreted in 
our EFT framework as the three-body eigenenergy, $-E=B_{\Lambda\Lambda}$, obtained as the likely ground-state solution to the homogeneous part of the integral equations. Additionally, in the cluster model 
framework it is conventional to define an {\it incremental binding energy} $\Delta B_{\Lambda\Lambda}$ which
is related to $B_{\Lambda\Lambda}$ (measured with respect to the $\Lambda\Lambda T$ three-particle breakup 
threshold) 
as~\cite{Filikhin:2002wm}
\begin{equation}
\Delta B_{\Lambda\Lambda}= B_{\Lambda\Lambda}-2{\mathcal B}^{avg}_{\Lambda}\,,
\label{eq:incr_binding}
\end{equation}
where,
\begin{equation}
{\mathcal B}^{avg}_{\Lambda}=\frac{1}{4} {\mathcal B}_{\Lambda}[0^+]+\frac{3}{4}{\mathcal B}_{\Lambda}[1^+]\,,
\label{eq:BLam_spin_avg}
\end{equation}
is the $(2J+1)$ {\it spin averaged} $\Lambda$-separation energy of the singlet and triplet two-body 
subsystems (interpreted in the EFT as the $(\Lambda T)_{s,t}$ subsystem averaged binding energy). Thus, 
the predicted values of $B_{\Lambda\Lambda}$ from past {\it ab initio} potential model analysis, such as in 
Ref.~\cite{Nemura:2004xb}, may be used to supplant the old results of $\Delta B_{\Lambda\Lambda}$ by 
reevaluating them using the recent experimental inputs for the $\Lambda$-separation energies of the ground 
(singlet) and first (triplet) excited states of the (${}_{\Lambda}^{4}{\rm H}\,,\,{}_{\Lambda}^{4}{\rm He}$) 
mirrors~\cite{Esser:2015trs,Schulz:2016kdc,Yamamoto:2015avw,Koike:2019rrs}.} ($B_{\Lambda\Lambda}$) for the 
mirror partners, predicted by the {\it ab initio} coupled channel potential model of Nemura 
{\it et al.}~\cite{Nemura:2004xb} using SVM analysis (cf. Table.~\ref{table:T2}\,). These predictions correspond to the 
two representative $S$-wave double-$\Lambda$ scattering lengths, namely, $a_{\Lambda\Lambda}=-0.91$ and 
$-1.37~{\rm fm}$, taken from the old Nijmegen hard-core potential models, mND$_S$ and ND$_S$, respectively, of 
Ref.~\cite{Nagels:1978sc}, but consistent with the constraints based on recent theoretical
analyses~\cite{Morita:2014kza,Ohnishi:2015cnu,Ohnishi:2016elb} based on RHIC data~\cite{Adamczyk:2014vca}.  

\vspace{0.1cm}

In Fig.~\ref{fig:fig3} we demonstrate the cutoff regulator dependence of the three-body coupling 
$g^{(A,B)}_3(\Lambda_c)$ for the $\Lambda\Lambda t$ system. The characteristic quasiperiodic cyclic singularities
reminiscent of the asymptotic limit cycle associated with the successive formation of three-body bound states is 
clearly evident in the nonasymptotic domain. Our finding in the three-body sector reveals good isospin symmetry 
between the two double-$\Lambda$-hypernuclear mirror partners with very little discernible difference in the RG 
behavior of each partner. Consequently, for brevity, we do not display the result for the $\Lambda\Lambda h$ 
system. As already pointed out, ideally the scale dependence of the type-A and type-B three-body couplings should be 
identical. However, owing to the small qualitative differences in rearrangements between the two types of elastic 
reaction channels where we only choose to introduce the counterterms (cf. Figs.~\ref{fig:fig1} and 
\ref{fig:fig2}\,), the type-B limit cycle plots are nominally shifted leftwards and downwards with respect to the
type-A limit cycle plots. In particular, due to considerable sensitivity to the small cutoff region, $\Lambda_c\lesssim 200$~MeV, 
the $N=0$ branch which is altogether washed out in the type-B plot, is still manifest in the type-A plot (top left
corner). However, this branch is not associated with the formation of an Efimov state. The ground 
($n=N-1=0$) state on the other hand is associated with the $N=1$ branch. Nevertheless, the regulator values, 
$\Lambda_c=(\Lambda_c)_N$, at which these couplings successively vanish remain unaltered in the two types of limit 
cycle plots. In each case the nonasymptotic RG limit cycle parameter $s_0$ can be calculated via the relation 
\begin{equation}
s_0=\frac{\pi}{\ln\left[\frac{(\Lambda_c)_{N+1}}{(\Lambda_c)_{N}}\right]}\,;  \quad N=1,2, \cdots
\end{equation} 
where $(\Lambda_c)_N$ is the momentum cutoff corresponding to the $N$th zero of $g^{(A,B)}_3$.
Using, say,  the $N=1,2$ values of $\Lambda_c$ we obtain $s_0=\pi/\ln[(\Lambda_c)_2/(\Lambda_c)_1]\approx 1.03$, which 
is nearly the same as the asymptotic values of $s^\infty_0$ given in Eq.~\eqref{eq:s0_asy}, irrespective of the chosen 
type of elastic channel. It is also notable that our $s_0$ or $s^\infty_0$ values agree well with typical values 
anticipated from the universal calibration curve for a mass imbalanced three-body system~\cite{Braaten:2004rn}, namely, 
the plot of exp$(\pi/s_0)$ versus the mass ratio $m_1/m_3$, with $m_1=m_2\equiv M_\Lambda$ and 
$m_3\equiv M_T\neq m_{1,2}$.

\vspace{0.1cm}

\begin{figure}[tbp]
    \centering
    \includegraphics[width=\columnwidth]{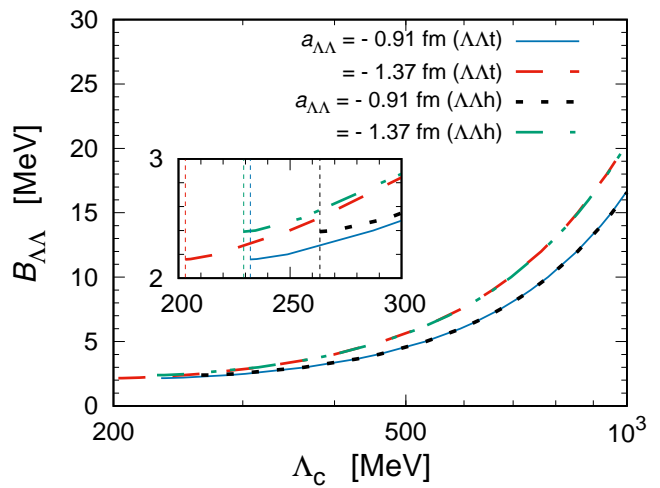}
    \caption{\label{fig:fig4} The cutoff regulator ($\Lambda_c$) dependence of the three-body 
             binding or the double-$\Lambda$-separation energy $B_{\Lambda\Lambda}$ (with respect to the 
             three-particle threshold) of $\Lambda\Lambda T$ mirror systems with the three-body
             couplings $g^{(A,B)}_3$ excluded. The plots correspond to the results for both 
             choices of the elastic channels. Two representative choices for the double-$\Lambda$ 
             scattering lengths are considered, namely, $a_{\Lambda\Lambda}=-0.91$~fm and 
             $-1.37$~fm, based on the old Nijmegen hard-core potential models, mND$_S$ and ND$_S$, 
             respectively~\cite{Nagels:1978sc}, and consistent with the recent theoretical 
             constraints~\cite{Morita:2014kza,Ohnishi:2015cnu,Ohnishi:2016elb} based on RHIC 
             data~\cite{Adamczyk:2014vca}. The vertical lines in the inset plot denote the 
             critical cutoffs, $\Lambda_c=\Lambda^{(n=0)}_{\rm crit}$, defined with 
             respect to the deeper particle-dimer thresholds, namely, the $\Lambda+u_0$ thresholds.
             Apart from the threshold regions, the results of both mirror partners are almost 
             identical.  } 
\end{figure}

Next we report on our regulator ($\Lambda_c$) dependence of $B_{\Lambda\Lambda}$ (cf. Fig.~\ref{fig:fig4}) obtained by 
numerically solving the homogeneous parts of the two sets of integral equations [cf. Eqs.~\eqref{eq:type-A} and 
\eqref{eq:type-B}\,], excluding the three-body contact interaction, i.e., $g^{(A,B)}_3\!\!=0$. Here we again consider the
two representative $S$-wave double-$\Lambda$ scattering lengths, namely, $a_{\Lambda\Lambda}=-0.91$~fm and 
$-1.37$~fm~\cite{Nagels:1978sc,Rijken:1998yy,Stoks:1999bz}, compatible with the range, 
$-1.92~{\rm fm} \lesssim a_{\Lambda\Lambda}\lesssim -0.5~{\rm fm}$~\cite{Morita:2014kza,Ohnishi:2015cnu,Ohnishi:2016elb}, 
constrained by RHIC data~\cite{Adamczyk:2014vca}. It may be noted that both choices (type-A and type-B) for the elastic 
channels yield identical cutoff dependence. Furthermore, both the double-$\Lambda$-hypernuclear mirror partners yield 
nearly identical results, apart from the expected ``mismatch" in threshold region (see inset plot of Fig.~\ref{fig:fig4}). 
Thus, it is interesting that despite the significant spin dependent {\it charge symmetry breaking} reflected in the 
two-body binding energies, e.g., $\delta {\mathcal B}_\Lambda[0^+] \gtrsim 200$~keV, the corresponding difference of the
spin-averaged binding energies, $\delta{\mathcal B}^{avg}_{\Lambda}\sim 5$~keV, is surprisingly small. This is easily 
seen using Eq.~\eqref{eq:BLam_spin_avg} with ${\mathcal B}^{avg}_{\Lambda}[{}_{\Lambda}^{4}{\rm H}]= 1.3395$~MeV and 
${\mathcal B}^{avg}_{\Lambda}[{}_{\Lambda}^{4}{\rm He}]= 1.3355$~MeV, based on the recent spectroscopic 
measurements~\cite{Yamamoto:2015avw,Esser:2015trs,Schulz:2016kdc,Koike:2019rrs} [cf. Table.~\ref{table:T3} and also 
Fig.~\ref{fig:fig0}]. Such a ``spin averaging" effect apparently gets implicitly reflected in the unrenormalized 
(regulator dependent) eigenenergies, $E(\Lambda_c)\equiv -B_{\Lambda\Lambda}$, obtained  via our integral equations 
with $g^{(A,B)}_3(\Lambda_c)=0$. The resulting difference of the double-$\Lambda$-separation energy ($B_{\Lambda\Lambda}$) 
between the (${}_{\Lambda\Lambda}^{\,\,\,\,5}{\rm H}\,,\,{}_{\Lambda\Lambda}^{\,\,\,\,5}{\rm He}$) mirror partners is 
evidently large, $\delta B_{\Lambda\Lambda}(\Lambda^{(n=0)}_{\rm crit})\gtrsim 200$ keV, around the particle-dimer 
thresholds, $\Lambda_c=\Lambda^{(n=0)}_{\rm crit}$ (i.e., the ground ($n=0$) state {\it critical cutoff} scales for the 
mirror partners\footnote{In our case in general, $\Lambda_c=\Lambda^{(n)}_{\rm crit}$, denotes the $n$th critical 
cutoff, defined as the cutoff scale at which the $n$th Efimov bound state emerges just above the {\it deeper} 
particle-dimer ($\Lambda+u_0$) breakup threshold $E={\mathcal E}^{thr}_{2(s)}$.}). However, this difference rapidly 
vanishes asymptotically ($\Lambda_c\to \infty$), ultimately leading to good charge and isospin symmetry. This feature will also
be apparent in our $B_{\Lambda\Lambda}\,$-$\,a_{\Lambda\Lambda}$ correlation results presented later in Table.~\ref{table:T5}. 
Notably, due to the absence of the three-body contact interactions to renormalize the integral equations, 
$B_{\Lambda\Lambda}$ is quite sensitive to the cutoff variations, which increase with increasing cutoff above the 
respective $\Lambda+u_0$ breakup thresholds. Moreover, it is apparent that the eigenenergies are also sensitive to the 
input double-$\Lambda$ scattering lengths, with $B_{\Lambda\Lambda}$ increasing with increasing $|a_{\Lambda\Lambda}|$. 
%--figure---------------------------------
%--figure---------------------------------
%%%%%%%%%%%%%%%%%%%%%%%%%%%%%%TABLE3%%%%%%%%%%%%%%%%%%%%%%%%%%%%%
\begin{table}[bp]
\begin{tabular}{ |c||c|c| }
\hline\hline
$\Lambda$-Hypernuclear & $\mathcal{B}_\Lambda[J^P]$ & $\gamma_{\Lambda T}\stackrel{!}{\leadsto}(2\mu_{\Lambda T} \mathcal{B}_\Lambda[J^P])^{1/2}$ \\
mirror states      &    (MeV) &  (MeV)\\ 
 \hline\hline
 ${}_{\,\,\,\,\Lambda}^{\,\,\,\,4}{\rm H}\,\,[0^+]$  & $2.157$~\cite{Esser:2015trs,Schulz:2016kdc}  & $\gamma_0\stackrel{!}{\leadsto}58.692$\\
 ${}_{\,\,\,\,\Lambda}^{\,\,\,\,4}{\rm H}\,\,(1^+)$  & $1.067$~\cite{Yamamoto:2015avw,Koike:2019rrs}& $\gamma_1\stackrel{!}{\leadsto}41.280$\\
\hline
 ${}_{\,\,\,\,\Lambda}^{\,\,\,\,4}{\rm He}\,\,[0^+]$ & $2.39$~\cite{Davis:2005mb}                   & $\gamma_0\stackrel{!}{\leadsto}61.779$\\
 ${}_{\,\,\,\,\Lambda}^{\,\,\,\,4}{\rm He}\,\,(1^+)$ & $0.984$~\cite{Yamamoto:2015avw,Koike:2019rrs}& $\gamma_1\stackrel{!}{\leadsto}39.641$\\
 \hline\hline
\end{tabular}
\caption{$\Lambda$-separation energies $\mathcal{B}_\Lambda[J^P=0^+,1^+]$ of the mirror states of 
        ($\!\!\!{}_{\,\,\,\,\Lambda}^{\,\,\,\,4}{\rm H}\,,\!\!\!{}_{\,\,\,\,\Lambda}^{\,\,\,\,4}{\rm He}$) 
        corresponding to the central values of the experimental results of 
        Refs.~\cite{Davis:2005mb,Yamamoto:2015avw,Esser:2015trs,Schulz:2016kdc,Koike:2019rrs} and 
        summarized in Fig.~\ref{fig:fig0}. In our EFT they are to be identified (``$\stackrel{!}{\leadsto}$" 
        denotes correspondence) with the particle-dimer breakup thresholds $-{\mathcal E}^{thr}_{2(s,t)}$ 
        for the $\Lambda\Lambda T$ systems or equivalently, the $u_{0,1}\equiv (\Lambda T)_{s,t}$ dimer 
        binding energies. The corresponding binding momenta $\gamma_{\Lambda T}\equiv \gamma_{0,1}$ are 
        inputs to our integral equations.}
\label{table:T3}
\end{table}
%%%%%%%%%%%%%%%%%%%%%%%%%%%%%%%%%%%%%%%%%%%%%%%%%%%%%%%%%%%%%%%%%

\vspace{0.1cm}

We emphasize that, although in our EFT framework the $u_{0,1}\equiv(\Lambda T)_{s,t}$ two-body subsystems introduce two 
relevant energy scales ${\mathcal E}^{thr}_{2(s,t)}$, it is the larger of the two particle-dimer thresholds, namely, the 
$\Lambda+u_0$ (singlet-dimer) threshold that is effectively associated with the formation of Efimov states. In fact,
irrespective of the chosen (type-A,B) elastic channels, our numerical evaluations of the integral equation only yield 
trimer states which are deeper than the $\Lambda+u_0$ thresholds, viz. $B_{\Lambda\Lambda}>{\mathcal B}_{\Lambda}[0^+]$ 
provided $\Lambda_c >\Lambda^{(n=0)}_{\rm crit}$. No numerically stable eigensolutions are obtained in the energy 
domain, ${\mathcal E}^{thr}_{2(s)}< E < {\mathcal E}^{thr}_{2(t)}$, lying in between the two thresholds. Thus, we should 
re-emphasize the correspondence of the $\Lambda\Lambda T \to \Lambda+u_0$ breakup threshold energies 
${\mathcal E}^{thr}_{2(s)}$ of the respective double-$\Lambda$-hypernuclear mirror partners to the $(\Lambda T)_s$ 
subsystem binding energies, vis-a-vis the $\Lambda$-separation energies ${\mathcal B}_{\Lambda}[0^+]$ of the ground
($J^P=0^+$) state of the 
$(\!\!\!{}_{\,\,\,\,\Lambda}^{\,\,\,\,4}{\rm H}\,,\!\!\!{}_{\,\,\,\,\Lambda}^{\,\,\,\,4}{\rm He})$ mirror partners, namely, 
\begin{eqnarray}
B_{\Lambda\Lambda}(\Lambda^{(n=0)}_{\rm crit})&\equiv&-{\mathcal E}^{thr}_{2(s)}=\frac{\gamma^2_0}{2\mu_{\Lambda T}}
\stackrel{!}{\leadsto}{\mathcal B}_{\Lambda}[0^+] 
\\
&=&
\begin{cases}
2.157~\text{MeV~\cite{Esser:2015trs,Yamamoto:2015avw,Schulz:2016kdc}} & \text{for}\,\,\, {}_{\,\,\,\,\Lambda}^{\,\,\,\,4}{\rm H}[0^+]\,, \\ 
2.39~\text{MeV~\cite{Davis:2005mb}}                  & \text{for}\,\,\, {}_{\,\,\,\,\Lambda}^{\,\,\,\,4}{\rm He}[0^+]\,.
\end{cases}
\nonumber
\label{breakup_threshold}
\end{eqnarray}
Here, the currently accepted central values of experimentally determined $\Lambda$-separation 
energies~\cite{Davis:2005mb,Yamamoto:2015avw,Esser:2015trs,Schulz:2016kdc,Koike:2019rrs} [cf. Table.~\ref{table:T3} 
and also Fig.~\ref{fig:fig0}] are used to fix the two-body input parameters of the $(\Lambda T)_{s,t}$ systems, namely, 
the binding momenta, defined by the correspondence, 
$\gamma_{0,1}\stackrel{!}{\leadsto}\sqrt{2\mu_{\Lambda T} \mathcal{B}_\Lambda[J^P=0^+,1^+]\,}$, which reflect the 
information regarding the two breakup thresholds in our integrals equations. These critical cutoffs for the ground 
($n=0$) states were tabulated earlier in Table.~\ref{table:T2}. The rightmost column in the same table also displays 
our cutoff values, $\Lambda_c=\Lambda^{(n=0)}_{\rm pot}$ that reproduce the double-$\Lambda$-separation energies 
$B_{\Lambda\Lambda}$ of Ref.~\cite{Nemura:2004xb}, begin interpreted as the plausible Efimov ground ($n=0$) state 
eigenenergies. Although  the $\Lambda^{(n=0)}_{\rm pot}$ values are significantly larger than the canonical hard scale 
of a ${}^{\pi\!\!\!/}$EFT, namely, $\Lambda_H\sim m_\pi$, they are nevertheless within a reasonable ballpark in context
of hypernuclear systems where one-pion exchanges are forbidden by virtue of isospin invariance. A more reasonable choice
of our EFT hard scale consistent with the low-energy symmetries in this case could be $\Lambda_H\gtrsim 2m_\pi$, with 
the $\Lambda$-$\Lambda$ interactions known to be dominated by $\pi\pi$ or the $\sigma$-{\it meson} exchange mechanism. It 
is, however, not inconceivable that a momentum scale of this magnitude is inconsistent with the $\Lambda\Lambda T$ bound
cluster {\it ansatz}, whereby the very existence of the core fields, $T\equiv t,h$ becomes questionable. 

%--figure---------------------------------
\begin{figure*}[t]
    \centering
    \includegraphics[width=\columnwidth]{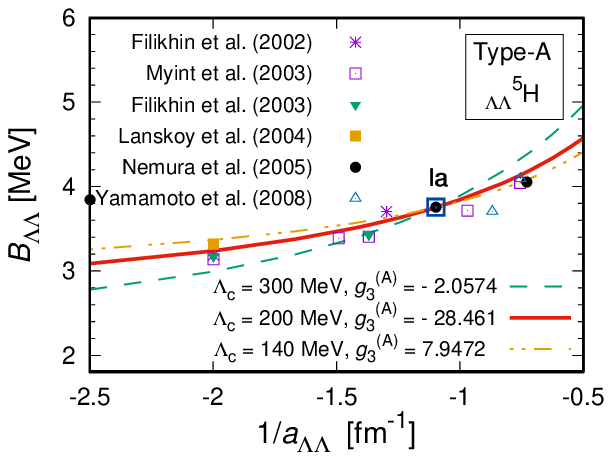}
    \includegraphics[width=\columnwidth]{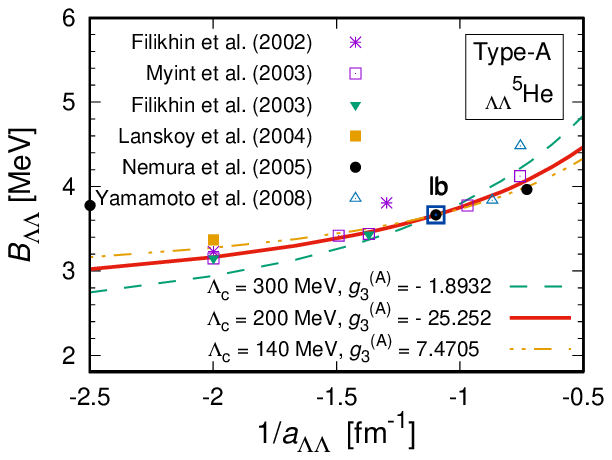}
    
    \vspace{-0.1cm}
    
    \caption{\label{fig:fig5} The double-$\Lambda$-separation energies $B_{\Lambda\Lambda}$ of 
             ${}_{\Lambda\Lambda}^{\,\,\,\,5}$H (left panel) and ${}_{\Lambda\Lambda}^{\,\,\,\,5}$He (right panel) 
             as a function of  the inverse of the $S$-wave double-$\Lambda$ scattering length $a^{-1}_{\Lambda\Lambda}$ 
             using different values of the three-body coupling $g^{(A)}_3$ at appropriate cutoff scales $\Lambda_c$. 
             These results correspond to the type-A choice of the elastic channel obtained using integral 
             equations~\eqref{eq:type-A}. The displayed data points correspond to our reevaluations [ via 
             Eq.~\eqref{eq:incr_binding}] of the past potential model-based predictions of  Refs.~\cite{Filikhin:2002wm,Filikhin:2003js,Myint:2002dp,Lanskoy:2003ia,Nemura:2004xb}  
             using the current experimental input for the $\Lambda$-separation energies 
             ${\mathcal B}_\Lambda[0^+,1^+]$ of 
             ($\!\!\!{}_{\,\,\,\,\Lambda}^{\,\,\,\,4}{\rm H}\,,\!\!\!{}_{\,\,\,\,\Lambda}^{\,\,\,\,4}{\rm He}$)~\cite{Yamamoto:2015avw,Esser:2015trs,Schulz:2016kdc,Koike:2019rrs}. In particular, the two data 
             points, namely, ``Ia": ($B_{\Lambda\Lambda}=3.750$~MeV, $a_{\Lambda\Lambda}=-0.91$~fm) for 
             ${}_{\Lambda\Lambda}^{\,\,\,\,5}$H and ``Ib": ($B_{\Lambda\Lambda}=3.660$~MeV, $a_{\Lambda\Lambda}=-0.91$~fm)
             for ${}_{\Lambda\Lambda}^{\,\,\,\,5}$He (large open squares), taken from Ref.~\cite{Nemura:2004xb} best serve
             to normalize our solutions to the integral equations. }
\end{figure*}
%--figure---------------------------------
%--figure---------------------------------
\begin{figure*}[tbp]
    \centering
    \includegraphics[width=\columnwidth]{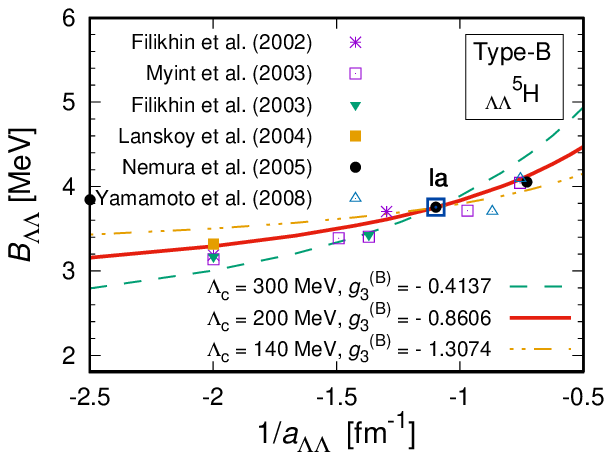}
    \includegraphics[width=\columnwidth]{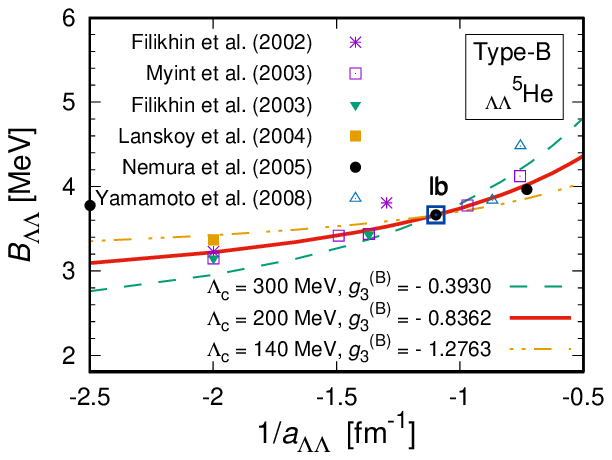}
    
    \vspace{-0.1cm}
    
    \caption{\label{fig:fig6} The double-$\Lambda$-separation energies $B_{\Lambda\Lambda}$ of 
             ${}_{\Lambda\Lambda}^{\,\,\,\,5}$H (left panel) and ${}_{\Lambda\Lambda}^{\,\,\,\,5}$He (right panel) 
             as a function of the inverse of the $S$-wave double-$\Lambda$ scattering length $a^{-1}_{\Lambda\Lambda}$ 
             using different values of the three-body coupling $g^{(B)}_3$ at appropriate cutoff scales $\Lambda_c$. 
             These results correspond to the type-B choice of the elastic channel obtained using integral 
             equations~\eqref{eq:type-B}. The displayed data points correspond to our reevaluations [ via 
             Eq.~\eqref{eq:incr_binding}] of the past potential model-based predictions of Refs.~\cite{Filikhin:2002wm,Filikhin:2003js,Myint:2002dp,Lanskoy:2003ia,Nemura:2004xb} 
             using the current experimental input for the $\Lambda$-separation energies 
             ${\mathcal B}_\Lambda[0^+,1^+]$ of 
             ($\!\!\!{}_{\,\,\,\,\Lambda}^{\,\,\,\,4}{\rm H}\,,\!\!\!{}_{\,\,\,\,\Lambda}^{\,\,\,\,4}{\rm 
             He}$)~\cite{Yamamoto:2015avw,Esser:2015trs,Schulz:2016kdc,Koike:2019rrs}. In particular, the two data 
             points, namely, ``Ia": ($B_{\Lambda\Lambda}=3.750$~MeV, $a_{\Lambda\Lambda}=-0.91$~fm) for 
             ${}_{\Lambda\Lambda}^{\,\,\,\,5}$H and ``Ib": ($B_{\Lambda\Lambda}=3.660$~MeV, $a_{\Lambda\Lambda}=-0.91$~fm) 
             for ${}_{\Lambda\Lambda}^{\,\,\,\,5}$He (large open squares), taken from Ref.~\cite{Nemura:2004xb} best serve 
             to normalize our solutions to the integral equations. }
\end{figure*}
%--figure---------------------------------

\vspace{0.1cm}

In Figs.~\ref{fig:fig5} and \ref{fig:fig6}, for each choice (type-A, -B) of the elastic channel, we plot our 
predictions for the $B_{\Lambda\Lambda}\,$-$\,a_{\Lambda\Lambda}$ correlation using different values of the three-body 
couplings $g^{(A,B)}_3$ at appropriate cutoff scales. Solutions to each set of integral equations [i.e., 
Eqs.~\eqref{eq:type-A} and \eqref{eq:type-B}] are normalized to a single (paired) data point which is conveniently
taken from the {\it ab initio} potential model analysis of Ref.~\cite{Nemura:2004xb}, each for 
${}_{\Lambda\Lambda}^{\,\,\,\,5}$H and ${}_{\Lambda\Lambda}^{\,\,\,\,5}$He, namely, the data points  ``Ia" 
($B_{\Lambda\Lambda}=3.750$~MeV, $a_{\Lambda\Lambda}=-0.91$~fm) and ``Ib" ($B_{\Lambda\Lambda}=3.660$~MeV, 
$a_{\Lambda\Lambda}=-0.91$~fm), respectively (cf. Table.~\ref{table:T2}\,). In particular, our EFT results for the 
choice of the regulator, $\Lambda_c=200$~MeV, corresponding to the three-body couplings, $g^{(A)}_3=-28.461$ and 
$g^{(B)}_3=-0.8606$ for ${}_{\Lambda\Lambda}^{\,\,\,\,5}$H, and $g^{(A)}_3=-25.252$ and $g^{(B)}_3=-0.8362$ for
${}_{\Lambda\Lambda}^{\,\,\,\,5}$He, agree reasonably well with the existing regulator independent potential model 
results~\cite{Filikhin:2002wm,Filikhin:2003js,Myint:2002dp,Lanskoy:2003ia,Nemura:2004xb} (provided of course that the original 
model predictions of $B_{\Lambda\Lambda}$ or $\Delta B_{\Lambda\Lambda}$, based on the superannuated 
${\mathcal B}_\Lambda[0^+,1^+]$ experimental data~\cite{Juric:1973zq,Davis:2005mb} are reevaluated using the current 
data~\cite{Yamamoto:2015avw,Esser:2015trs,Schulz:2016kdc,Koike:2019rrs}).\footnote{The past potential model analyses 
used different three-body techniques to determine either $B_{\Lambda\Lambda}$ or $\Delta B_{\Lambda\Lambda}$ in one 
of two ways: 
(i) {\it ab initio} determination, using elementary two- and three-body baryonic interactions, and 
(ii) cluster model determination, relying on the elementary four-body inputs (or equivalently, the 
two-body inputs in our particle-dimer cluster scenario), namely, the $\Lambda$-separation energies 
${\mathcal B}_\Lambda[0^+,1^+]$ of 
($\!\!\!{}_{\,\,\,\,\Lambda}^{\,\,\,\,4}{\rm H}\,,\!\!\!{}_{\,\,\,\,\Lambda}^{\,\,\,\,4}{\rm He}$) from old emulsion
studies~\cite{Juric:1973zq,Davis:2005mb}. 
With the advent of the recent high-precision data on ${\mathcal B}_\Lambda[0^+,1^+]$ from MAMI and
J-PARC~\cite{Yamamoto:2015avw,Esser:2015trs,Schulz:2016kdc,Koike:2019rrs}, the old emulsion works have now been 
superseded. Consequently, all model data points displayed in Figs.~\ref{fig:fig5} and \ref{fig:fig6} correspond to 
our reevaluated $B_{\Lambda\Lambda}$ values from the old $\Delta B_{\Lambda\Lambda}$ model results using the current 
data {\em via} Eq.~\eqref{eq:incr_binding}. There is, however, a caveat to these figures: in the absence of updated 
results of the old cluster model analyses~\cite{Filikhin:2002wm,Filikhin:2003js,Myint:2002dp,Lanskoy:2003ia}, it is 
likely that some of the our reevaluated $B_{\Lambda\Lambda}$ ``model data points" may be nominally faulty in using
the old model $\Delta B_{\Lambda\Lambda}$ inputs, owing to certain degree of residual dependence on the 
superannuated ${\mathcal B}_{\Lambda}[0^+,1^+]$ data. } To this end, each solid red curve in the 
figures represents our EFT generated {\it calibration curve} reflecting the inherent nature of the
$B_{\Lambda\Lambda}\,$-$\,a_{\Lambda\Lambda}$ correlations of the $\Lambda\Lambda T$ mirror systems. Thus, in the remaining
part of our analysis we use the correlation plots corresponding to $\Lambda_c=200$~MeV to predict $B_{\Lambda\Lambda}$
for arbitrary values of $a_{\Lambda\Lambda}$.      

%--figure---------------------------------
\begin{figure*}[tbp]
    \centering
    \includegraphics[width=\columnwidth]{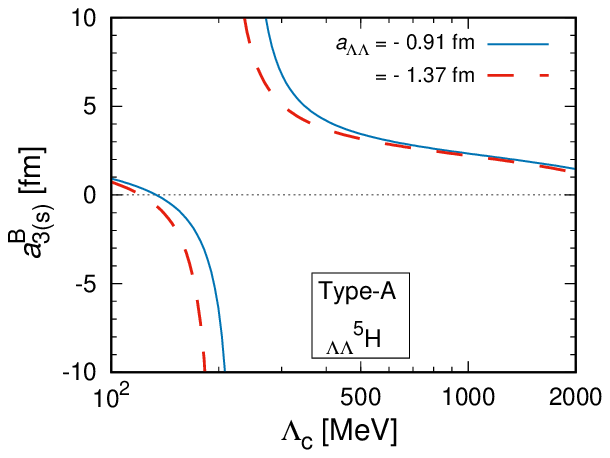}
    \includegraphics[width=\columnwidth]{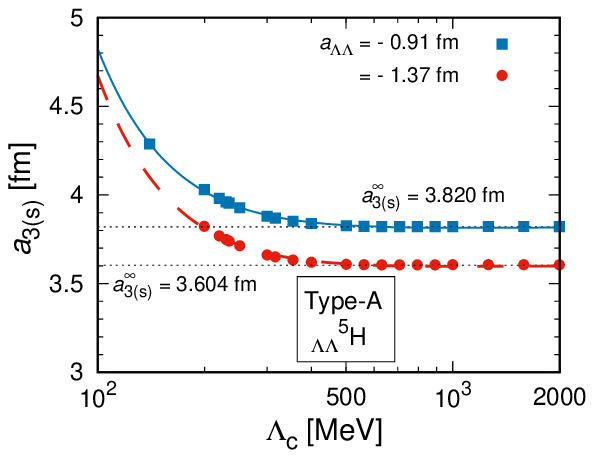}
    
    \vspace{-0.1cm}
    
    \caption{\label{fig:fig7} The EFT predicted regulator ($\Lambda_c$) dependence of the $J=1/2$ $S$-wave $\Lambda\,$-$\,(\Lambda t)_s$ 
             scattering length $a_{3(s)}$ for the ${}_{\Lambda}^{4}{\rm H}[0^+]\,$-$\,\Lambda$ scattering without (left panel) and 
             with (right panel) the three-body coupling $g^{(A)}_3$. Two representative values of the Nijmegen hard-core potential
             model extracted double-$\Lambda$ scattering lengths are used, namely, 
             $a_{\Lambda\Lambda}=-0.91,\, -1.37$~fm~\cite{Nagels:1978sc}, which are consistent with recent RHIC data
             analyses~\cite{Morita:2014kza,Ohnishi:2015cnu,Ohnishi:2016elb}. The input double-$\Lambda$-separation energies
             $B_{\Lambda\Lambda}$ needed to fix $g^{(A)}_3(\Lambda_c)$ for renormalization are obtained by using our EFT calibration
             curves (solid red line in Fig.~\ref{fig:fig5}; see also Table.~\ref{table:T5}\,). The unrenormalized (bare) 
             scattering length is denoted $a^B_{3(s)}$. The smooth curves in the right panel represent fits to the data points 
             based on the power series ansatz, Eq.~\eqref{eq:a3_fit}. The corresponding results for $\Lambda\Lambda h$ or
             ${}_{\Lambda}^{4}{\rm He}[0^+]\,$-$\,\Lambda$ scattering being very similar, are not displayed.}
\end{figure*}
%-----------------------------------------
%-----------------------------------------
\begin{figure*}[tbp]
    \centering
    \includegraphics[width=\columnwidth]{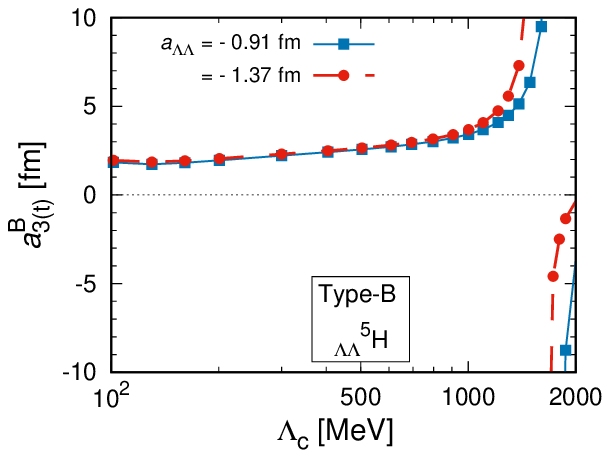}
    \includegraphics[width=\columnwidth]{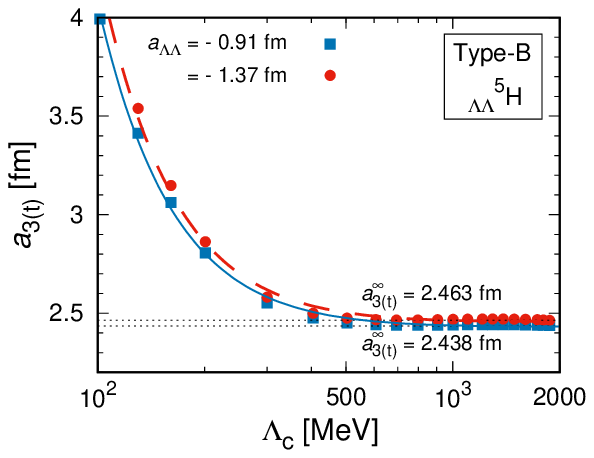}
    
    \vspace{-0.1cm}
    
    \caption{\label{fig:fig8} The EFT predicted regulator ($\Lambda_c$) dependence of the $J=1/2$ $S$-wave $\Lambda\,$-$\,(\Lambda t)_t$ 
             scattering length $a_{3(t)}$ for the ${}_{\Lambda}^{4}{\rm H}[1^+]\,$-$\,\Lambda$ scattering without (left panel) and 
             with (right panel) the three-body coupling $g^{(B)}_3$. Two representative values of the Nijmegen hard-core potential
             model extracted double-$\Lambda$ scattering lengths are used, namely, 
             $a_{\Lambda\Lambda}=-0.91,\, -1.37$~fm~\cite{Nagels:1978sc}, which are consistent with recent RHIC data 
             analyses~\cite{Morita:2014kza,Ohnishi:2015cnu,Ohnishi:2016elb}. The input double-$\Lambda$-separation energies 
             $B_{\Lambda\Lambda}$ needed to fix $g^{(B)}_3(\Lambda_c)$ for renormalization are obtained by using our EFT calibration 
             curves (solid red line in Fig.~\ref{fig:fig6}; see also Table.~\ref{table:T5}\,). The unrenormalized (bare) 
             scattering length is denoted $a^B_{3(t)}$. The smooth curves in the right panel represent fits to the data points 
             based on the power series ansatz, Eq.~\eqref{eq:a3_fit}. The corresponding results for $\Lambda\Lambda h$ or 
             ${}_{\Lambda}^{4}{\rm He}[1^+]\,$-$\,\Lambda$ scattering being very similar, are not displayed.}
\end{figure*}
%-----------------------------------------

\vspace{0.1cm}

The final part of our EFT analysis is concerned with the preliminary estimation of the $S$-wave three-body
scattering lengths $a_{\Lambda\Lambda T}$, namely, the ${}_{\Lambda}^{4}{\rm H}\,$-$\,\Lambda$ and  
${}_{\Lambda}^{4}{\rm He}\,$-$\,\Lambda$ scattering lengths. For this purpose, we numerically solve the 
two sets of coupled integral equations for the renormalized on-shell elastic $K$-matrix elements 
${\mathbb K}^{A,B}_a(k,k)$ in each case [cf. Eqs.~\eqref{eq:type-KA} and \eqref{eq:type-KB}], which yield 
the scattering lengths in the threshold limit ($k\to 0$). Care must be taken to bypass the poles of the 
dimer propagators originating in the kinematical scattering domain close to the respective particle-dimer 
thresholds. In this regard, we have implemented a numerical methodology of solving a multidimensional 
generalization of principal value prescription modified integral equations, originally developed by 
Kowalski and Noyes~\cite{Kowalski,Noyes} (see also Ref.~\cite{Gloeckle}) for the one-dimensional case.  

\vspace{0.1cm}

Figures~\ref{fig:fig7} and \ref{fig:fig8} display the cutoff scale dependence of the $\Lambda\,$-\,$(\Lambda t)_{s,t}$ 
scattering lengths for the ${}_{\Lambda}^{4}{\rm H}[0^+,1^+]\,$-$\,\Lambda$ scattering processes for the two input 
double-$\Lambda$ scattering lengths, namely, $a_{\Lambda\Lambda}=-0.91$~fm and $-1.37$~fm~\cite{Nagels:1978sc}. In 
this case the results for the $\Lambda\Lambda T$ mirror partners are imperceptibly close to each other, so that we 
graphically display the results for only one of them, say, the $\Lambda\Lambda t$ system, although a consolidated 
summary of our numerical predictions for both mirror partners are tabulated in Table.~\ref{table:T5}. It is, however,
worth mentioning that in contrast with our universal LO EFT prediction with very little observable difference between 
the $\Lambda\Lambda T$ mirrors, somewhat large isospin-breaking corrections have been reported for these systems in 
the context of existing potential-model analyses. This leads to significant differences in the model predictions of the two- and
three-body binding energies~\cite{Lanskoy:2003ia,Gal:2015bfa}. Such precision effects are not captured without a 
subleading-order EFT calculation, which is beyond the present scope.

\vspace{0.05cm}

In contrast with little or no quantitative difference in results corresponding to each choice (type-A, -B) of the bound-state solutions, significant qualitative differences arise in the respective scattering domains.
In the left panel plots of Figs.~\ref{fig:fig7} and \ref{fig:fig8}, which exclude the three-body contact interactions, 
the unregulated (bare) scattering amplitudes $a^{\rm B}_{3(s,t)}$ depend sensitively on the cutoff scale and diverge for 
specific values of $\Lambda_c$ associated with the successive emergence of three-body bound states. Moreover for $\Lambda_c\lesssim 200$~MeV, the very first pole-like feature seen in the 
unrenormalized type-A amplitude in Fig.~\ref{fig:fig7}, is missing from the unrenormalized type-B amplitude displayed in 
Fig.~\ref{fig:fig8}. This feature is concomitant with the associated limit cycle behavior of the three-body 
systems (cf. Fig.\ref{fig:fig3}) where the $N=0$ branch is found to be altogether missing in the type-B plots. 
Nevertheless, such unphysical singularities in the scattering amplitude are renormalized by the introduction of the 
scale dependent couplings $g^{(A,B)}_3(\Lambda_c)$, as revealed in the right panel plots which are free of singularities.
We find that the renormalized scattering lengths $a_{3(s,t)}$ smoothly decreases with increasing $\Lambda_c$ converging 
asymptotically for $\Lambda_c\gtrsim 500$ MeV. These asymptotic values $a_{3(s,t)}(\Lambda_c\to \infty)$ precisely yield 
our EFT predictions of the three-body scattering lengths corresponding to type-A,B choices of the elastic channels. To obtain the asymptotic values $a^\infty_{3(s,t)}$, we use the power series fitting ansatz for small ${\bar Q}/\Lambda_c$, namely,
\begin{eqnarray}
a_{3(s,t)}(\Lambda_c)=a^\infty_{3(s,t)} \!\! \left[1\!+\!\alpha_{s,t}\!\left(\frac{\bar Q}{\Lambda_c}\right)
\!+\!\beta_{s,t} \!\left(\frac{\bar Q}{\Lambda_c}\right)^2\!\!+\cdots\! \right],\,\,\,\quad
\label{eq:a3_fit}
\end{eqnarray}
applied to our generated data points, as shown in the figures, obtained as solutions to the renormalized $K$-matrix 
equations~\eqref{eq:type-KA} and \eqref{eq:type-KB}. As estimated earlier, ${\bar Q}\sim 50$~MeV can be conveniently 
taken as the generic momentum scale of the underlying dynamics, with $\alpha_{s,t},\, \beta_{s,t}$ and $a^\infty_{3(s,t)}$ 
being the fitting parameters. Thus, the corresponding fitting curves, as displayed Figs.~\ref{fig:fig7} and \ref{fig:fig8} 
(right panel plots), yield the respective three-body scattering lengths by extrapolating to $\Lambda_c\to \infty$. We note
that a similar {\it ansatz} was recently used in the SVM ${}^{\pi\!\!\!/}$EFT calculation of Ref.~\cite{Contessi:2019csf} 
to estimate the $\Lambda$-separation energies ${\mathcal B}_{\Lambda}$ of the 
(${}_{\Lambda\Lambda}^{\,\,\,\,5}{\rm H}\,,\,{}_{\Lambda\Lambda}^{\,\,\,\,5}{\rm He}$) mirror partners. 

\vspace{0.1cm}

Table.~\ref{table:T5} summarizes our numerical estimates of the $S$-wave renormalized type-A, -B three-body scattering lengths 
$a_{3(s,t)}\equiv a_{3(s,t)}(\Lambda_c\to \infty)$, as well as the {\it spin-averaged} values $a_{\Lambda\Lambda T}$ for 
different $a_{\Lambda\Lambda}$ inputs within the current theoretically feasible range, 
$-1.92\,\, {\rm fm} \lesssim a_{\Lambda\Lambda} \lesssim -0.5 \,\, {\rm fm}$, based on RHIC data
analyses~\cite{Morita:2014kza,Ohnishi:2015cnu,Ohnishi:2016elb}.\footnote{In contrast, the same RHIC data previously analyzed 
by the STAR Collaboration~\cite{Adamczyk:2014vca} suggested a positive value of the scattering length. It is notable, 
however, that our analysis in this work is {\it only} justified on the basis of a virtual bound $\Lambda\Lambda$ state. Hence,
we restrict our analysis to negative $a_{\Lambda\Lambda}$ values only.} The chosen $a_{\Lambda\Lambda}$ values range from 
those extracted from the old Nijmegen potential models
(e.g., NHC-F, \,NSC97e, \,ND, \,ND$_S$, \,mND$_S$)~\cite{Nagels:1978sc,Rijken:1998yy,Stoks:1999bz}, including more recent ones 
based on dispersion techniques~\cite{Gasparyan:2011kg} and RHIC thermal correlation model 
analysis~\cite{Morita:2014kza,Ohnishi:2015cnu,Ohnishi:2016elb}, and up to the most recent ones based on lattice simulations by 
the HAL QCD Collaboration~\cite{Sasaki:2019qnh}. In particular, a representative value of $a_{\Lambda\Lambda}=-0.8$~fm was 
suggested in the recent SVM ${}^{\pi\!\!\!/}$EFT calculation~\cite{Contessi:2019csf}, using which ${\mathcal B}_{\Lambda}$ 
of ${}_{\Lambda\Lambda}^{\,\,\,\,5}{\rm H}$ was predicted to be about $1.14$~MeV. This value can be compared with our 
estimation displayed in Table.~\ref{table:T6}, which in our $\Lambda\Lambda t$ cluster scenario may be naively obtained as
%%%%%%%%%%%%%%%%%%%%%%%%%%%%%%%%%%%%%%%%%%TABLE3%%%%%%%%%%%%%%%%%%%%%%%%%%%%%%%%%%%%
\begin{table*}[tbp]
\begin{tabular}{ |c|c||c|c||c|c||c| }
\hline\hline
Hypernucleus & Scattering length & Type-A & Type-A & Type-B & Type-B & $(2J+1)$ average   \\
   $(J=\frac{1}{2})$&  $a_{\Lambda\Lambda}$ (fm) 
   &$B_{\Lambda\Lambda}$ (MeV)&$a_{3(s)}(\Lambda_c\to \infty)$ (fm)&$B_{\Lambda\Lambda}$ (MeV)& $a_{3(t)}(\Lambda_c\to \infty)$ (fm)& $a_{\Lambda\Lambda T}$ (fm)\\
\hline\hline
                                          & -0.50 (NSC97e)~\cite{Rijken:1998yy,Stoks:1999bz}    & 3.236 & 4.258 & 3.292 & 2.388 & 2.968 \\
                                          & -0.60 (DR)~\cite{Gasparyan:2011kg}                  & 3.377 &4.109 & 3.418 & 2.404 & 2.925  \\
                                          & -0.73 (NHC-F)~\cite{Nagels:1978sc}                  & 3.544 & 3.964 & 3.567 & 2.420 & 2.885 \\ 
                                          & -0.77 (ND)~\cite{Rijken:1998yy,Stoks:1999bz}        & 3.592 & 3.927 & 3.610 & 2.425 & 2.875 \\
                                          & -0.80 (SVM)~\cite{Contessi:2019csf}& 3.627 & 3.902  & 3.641 & 2.428 & 2.868 \\
                                          & -0.81 (HAL QCD)~\cite{Sasaki:2019qnh}               & 3.639 & 3.894 & 3.651 & 2.429 & 2.866 \\    \cline{2-7} 
${}_{\Lambda\Lambda}^{\,\,\,\,5}{\rm H}$  & {\bf -0.91} (mND$_S$)~\cite{Nagels:1978sc} & 
                                            {\bf 3.750}~\cite{Nemura:2004xb} & 3.821   & 
                                            {\bf 3.750}~\cite{Nemura:2004xb} & 2.438   &   2.847 \\ \cline{2-7} 
                                          & -1.20 (DR)~\cite{Gasparyan:2011kg}                  & 4.030 & 3.668 & 3.997 & 2.456 & 2.809 \\
                                          & -1.25 (RHIC)~\cite{Morita:2014kza,Ohnishi:2015cnu,Ohnishi:2016elb}  & 4.073 & 3.648 & 4.034 &   2.459 & 2.804 \\
                                          & -1.32 (NHC-F)~\cite{Nagels:1978sc}                  & 4.131 & 3.622 & 4.085 & 2.462 & 2.797 \\ 
                                          & -1.37 (ND$_S$)~\cite{Nagels:1978sc}                 & 4.170 & 3.605 & 4.119 & 2.463 & 2.793 \\
                                          & -1.80 (DR)~\cite{Gasparyan:2011kg}                  & 4.461 & 3.493 & 4.374 & 2.473 & 2.764 \\
                                          & -1.92 (RHIC)~\cite{Morita:2014kza,Ohnishi:2015cnu,Ohnishi:2016elb} & 4.530   & 3.470 & 4.434 &   2.474 & 2.757  \\       
\hline\hline
 					                      & -0.50 (NSC97e)~\cite{Rijken:1998yy,Stoks:1999bz}    & 3.163 & 4.714 & 3.221 & 1.831 & 2.841 \\
                                          & -0.60 (DR)~\cite{Gasparyan:2011kg}                  & 3.298 & 4.461 & 3.341 & 1.837 & 2.740 \\
                                          & -0.73 (NHC-F)~\cite{Nagels:1978sc}                  & 3.460 & 4.229 & 3.484 & 1.843 & 2.649 \\ 
                                          & -0.77 (ND)~\cite{Rijken:1998yy,Stoks:1999bz}        & 3.506 & 4.173 & 3.525 & 1.845 & 2.628 \\
                                          & -0.80 (SVM)~\cite{Contessi:2019csf}& 3.541 & 4.134  & 3.555 & 1.846 & 2.613 \\
                                          & -0.81 (HAL QCD)~\cite{Sasaki:2019qnh}               & 3.552 & 4.121 & 3.565 & 1.846 & 2.608 \\ 
                                            \cline{2-7}
${}_{\Lambda\Lambda}^{\,\,\,\,5}{\rm He}$ & {\bf -0.91} (mND$_S$)~\cite{Nagels:1978sc}  & 
                                            {\bf 3.660}~\cite{Nemura:2004xb} & 4.012    & 
                                            {\bf 3.660}~\cite{Nemura:2004xb} & 1.849    &   2.567 \\ \cline{2-7}  
                                          & -1.20 (DR)~\cite{Gasparyan:2011kg}                  & 3.934 & 3.793 & 3.899 & 1.853 & 2.485 \\
                                          & -1.25 (RHIC)~\cite{Morita:2014kza,Ohnishi:2015cnu,Ohnishi:2016elb} & 3.976 & 3.766 & 3.935 & 1.854 & 2.474 \\
                                          & -1.32(NHC-F)~\cite{Nagels:1978sc}                   & 4.032 & 3.730 & 3.984 & 1.854 & 2.461 \\
                                          & -1.37(ND$_S$)~\cite{Nagels:1978sc}                   &4.071 & 3.707 & 4.018 & 1.854 & 2.452 \\
                                          & -1.80 (DR)~\cite{Gasparyan:2011kg}                  & 4.357 & 3.558 & 4.266 & 1.853 & 2.396 \\
                                          & -1.92 (RHIC)~\cite{Morita:2014kza,Ohnishi:2015cnu,Ohnishi:2016elb} & 4.425   & 3.528 &  4.324 & 1.852 & 2.384   \\            
\hline\hline 
\end{tabular}
\caption{The EFT predicted $J=1/2$ $S$-wave $\Lambda\Lambda T$ scattering lengths $a_{\Lambda\Lambda T}$ 
         [cf. Eq.~\eqref{eq:aLL}\,] of the double-$\Lambda$-hypernuclear mirror partners 
         (${}_{\Lambda\Lambda}^{\,\,\,\,5}{\rm H}\,,\,{}_{\Lambda\Lambda}^{\,\,\,\,5}{\rm He}$), obtained 
         for the central values of the $S$-wave scattering length $a_{\Lambda\Lambda}$ based on various 
         phenomenological analyses, e.g., old Nijmegen potential models (e.g., 
         NHC-F,\,NSC97e,\,ND,\,ND$_S$,\,mND$_S$)~\cite{Nagels:1978sc,Rijken:1998yy,Stoks:1999bz}, 
         dispersion relations (DR)~\cite{Gasparyan:2011kg}, thermal correlation model of relativistic 
         heavy-ion collisions (RHIC)~\cite{Morita:2014kza,Ohnishi:2015cnu,Ohnishi:2016elb}, {\it ab initio} 
         ${}^{\pi\!\!\!/}$EFT (SVM)~\cite{Contessi:2019csf}, and lattice QCD (HAL QCD)~\cite{Sasaki:2019qnh}, 
         consistent with the currently accepted range, 
         $-1.92\,\, {\rm fm} \lesssim a_{\Lambda\Lambda} \lesssim -0.5 \,\, 
         {\rm fm}$~\cite{Morita:2014kza,Ohnishi:2015cnu,Ohnishi:2016elb}. All the displayed 
         double-$\Lambda$-separation energies $B_{\Lambda\Lambda}$, excepting the two normalization values 
         taken from the potential model {\it ab initio} SVM analysis of Ref.~\cite{Nemura:2004xb} (shown in 
         bold), are obtained using our calibration curves for the choice of the cutoff scale, 
         $\Lambda_c=200$~MeV.}
\label{table:T5}
\end{table*}
%%%%%%%%%%%%%%%%%%%%%%%%%%%%%%%%%%%%%%%%%%%%%%%%%%%%%%%%%%%%%%%%%%%%
\begin{eqnarray}
{\mathcal B}_{\Lambda}\left({}_{\Lambda\Lambda}^{\,\,\,\,5}{\rm H}\right)
&=&\Delta B_{\Lambda\Lambda}\left({}_{\Lambda\Lambda}^{\,\,\,\,5}{\rm H}\right)
+{\mathcal B}^{avg}_{\Lambda}\left(\!\!\!{}_{\,\,\,\,\Lambda}^{\,\,\,\,4}{\rm H}\right)
\nonumber\\
&=& B_{\Lambda\Lambda}({\rm Avg})-{\mathcal B}^{avg}_{\Lambda}(\!\!\!{}_{\,\,\,\,\Lambda}^{\,\,\,\,4}{\rm H})\,,
\label{Lambda_SE}
\end{eqnarray}
where 
\begin{eqnarray}
B_{\Lambda\Lambda}({\rm Avg})=\frac{1}{2}\Big[B_{\Lambda\Lambda}(\text{type-A})+B_{\Lambda\Lambda}(\text{type-B})\Big]
\end{eqnarray}
%%%%%%%%%%%%%%%%%%%%%%%%%%%%%%%%%%%%%%%%%%%%%%%%%%%%%%%%%%%%%%%%%%%%
\begin{table}[tbp]
\begin{tabular}{ |c||c|c| }
\hline\hline
Hypernucleus      & This work, Eq.~\eqref{Lambda_SE} & Ref.~\cite{Contessi:2019csf}  \\
$(J=\frac{1}{2},\,I=\frac{1}{2})$ & ${\mathcal B}_{\Lambda}$ (MeV)    & ${\mathcal B}_{\Lambda}$ (MeV)  \\
\hline\hline
${}_{\Lambda\Lambda}^{\,\,\,\,5}{\rm H}$   &  2.295                & 1.14$\pm0.01_{-0.26}^{+0.44}$  \\
\hline
 ${}_{\Lambda\Lambda}^{\,\,\,\,5}{\rm He}$ &  2.212 &  - \\

\hline\hline 
\end{tabular}
\caption{The $\Lambda$-separation energies, namely, ${\mathcal B}_{\Lambda}({}_{\Lambda\Lambda}^{\,\,\,\,5}{\rm H})$ 
         and ${\mathcal B}_{\Lambda}({}_{\Lambda\Lambda}^{\,\,\,\,5}{\rm He})$, corresponding the representative value,  
         $a_{\Lambda\Lambda}=-0.80$ fm. The result for ${}_{\Lambda\Lambda}^{\,\,\,\,5}{\rm H}$ of
         Ref.~\cite{Contessi:2019csf} is displayed for comparison.}
\label{table:T6}
\end{table}
%%%%%%%%%%%%%%%%%%%%%%%%%%%%%%%%%%%%%%%%%%%%%%%%%%%%%%%%%%%%%%%%%%%%
is the ordinary mean of the type-A and type-B double-$\Lambda$-separation energies of 
${}_{\Lambda\Lambda}^{\,\,\,\,5}{\rm H}$ obtained from Table.~\ref{table:T5}. Likewise, we also obtain the estimate for 
${\mathcal B}_{\Lambda}\left({}_{\Lambda\Lambda}^{\,\,\,\,5}{\rm He}\right)$, as displayed in Table.~\ref{table:T6}.
Consequently, the difference of the two $\Lambda$-separation energies, namely, 
${\mathcal B}_{\Lambda}({}_{\Lambda\Lambda}^{\,\,\,\,5}{\rm He})
-{\mathcal B}_{\Lambda}({}_{\Lambda\Lambda}^{\,\,\,\,5}{\rm H})=9$~keV,  
yields a naive estimate of the charge-symmetry-breaking effects inherent to these double-$\Lambda$-hypernuclei. This 
is indeed small in comparison with that in the two-body sector where $\delta {\mathcal B}_\Lambda[0^+]\sim 200$~MeV
and $\delta {\mathcal B}_\Lambda[1^+]\sim 100$~MeV. However, the charge asymmetry noted here does not directly reflect 
anything regarding the underlying low-energy EFT dynamics with no isospin breaking terms included in the effective 
Lagrangian at LO, but rather a consequence of using physical masses and phenomenologically fixed inputs. Nevertheless, 
given the broad range of acceptable input values of the $S$-wave double-$\Lambda$ scattering lengths, namely, with 
$\delta a_{\Lambda\Lambda}\approx 1.42$~fm, the corresponding variations in the spin-averaged three-body scattering 
lengths turn out to be quite nominal, i.e., $\delta a_{\Lambda\Lambda T}\lesssim 0.4$~fm. But, it may be noticed that 
in contrast particularly the type-A scattering lengths $a_{3(s)}$ exhibit significant variations depending on the 
$(a_{\Lambda\Lambda},\, B_{\Lambda\Lambda})$ inputs. In fact the behavior, of $a_{3(s)}$ and $a_{3(t)}$ are turn out to be
quite the opposite, with the former increasing and the latter decreasing with both $|a_{\Lambda\Lambda}|$ and 
$B_{\Lambda\Lambda}$ decreasing. 

\vspace{0.1cm}

\begin{figure*}
    \centering
    \includegraphics[width=\columnwidth]{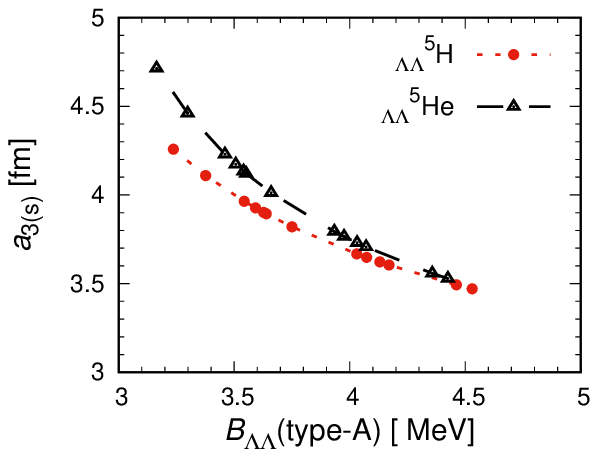} 
    \includegraphics[width=\columnwidth]{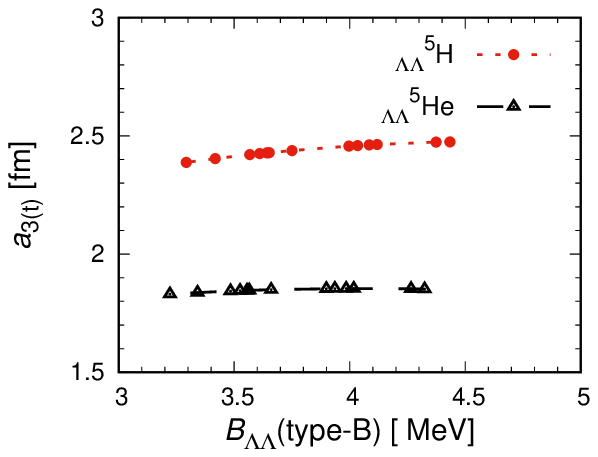}
    
    \vspace{0.3cm}
    
    \includegraphics[width=\columnwidth]{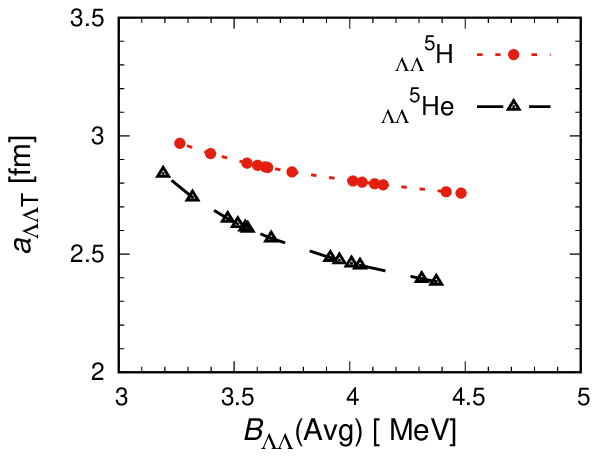}
    \caption{\label{fig:fig9} Phillips-lines for the type-A elastic channel, i.e., ${}_{\Lambda}^{4}{\rm H}[0^+]\,$-$\,\Lambda$ 
             and ${}_{\Lambda}^{4}{\rm He}[0^+]\,$-$\,\Lambda$ scatterings (upper left panel) and the type-B elastic channel, i.e.,
             ${}_{\Lambda}^{4}{\rm H}[1^+]\,$-$\,\Lambda$ and ${}_{\Lambda}^{4}{\rm He}[1^+]\,$-$\,\Lambda$ scatterings (upper 
             right panel) are displayed. The lower panel displays the ``physical" Phillips-lines corresponding to the spin-averaged 
             scattering lengths $a_{\Lambda\Lambda T}$ plotted as a function the mean values of the three-body binding energy,
             namely, $B_{\Lambda\Lambda}({\rm Avg})=\frac{1}{2}\left[B_{\Lambda\Lambda}(\text{type-A})
             +B_{\Lambda\Lambda}(\text{type-B})\right]$, obtained from Table.~\ref{table:T5}.}
\end{figure*}

The above discussed features are depicted clearly in the Phillips-line plots~\cite{Phillips:1968zze} shown in upper the
panels of Fig.~\ref{fig:fig9}, for each choice (type-A, -B) of the elastic channel. Interestingly, the
variation of the spin-averaged scattering length $a_{\Lambda\Lambda T}$, in what may be termed
as the ``physical" Phillips plot (lower panel) turns out to be significantly moderate. Evidently, the type-A Phillips
plots are in accordance with the expected behavior of the three-body binding energies varying inversely as the 
three-body scattering lengths, accounting for their characteristic negative slopes. In contrast, the observed positive
slope of the type-B Phillips plot may seem rather counterintuitive. It is noteworthy that these contrasting type-A, -B 
results are neither dependent on the nature of the three-body contact interactions used nor any artifact of the 
renormalization methods adopted in each case [cf. discussion below Eq.~\eqref{eq:g3_modified_K}]. This is easily 
understood by comparing the plots for the unrenormalized type-A, -B scattering lengths, which exhibit the same contrasting
features. In this context, we also note that in determining $a_{3(s)}$, only the dynamics near the deeper threshold, 
namely, the particle-dimer $\Lambda+u_0$ threshold, is relevant, whereas the dynamics of both thresholds ($\Lambda+u_{0,1}$)
contribute in the determination of $a_{3(t)}$. Although an unambiguous physical reasoning behind this contrasting 
behavior could not be ascertained, a plausible explanation may be attributed to the underlying nature of the off-shell 
dynamics arising due to the complex interplay between the two thresholds.

\vspace{0.1cm}

To test this hypothesis we took the strategy of considering a hypothetical (unphysical) scenario in which the triplet 
and singlet $\Lambda T$ subsystems are completely decoupled from each other to avoid the simultaneous contribution of both the 
particle-dimer thresholds for each $\Lambda\Lambda T$ systems. In other words, this is essentially 
tantamount to the removal of the triplet-dimer field $u_1$ contributions in the type-A integral 
equations~\eqref{eq:type-A}, and the singlet-dimer field $u_0$ contributions in the type-B integral 
equations~\eqref{eq:type-B}. The resulting $\Lambda\Lambda T$ dynamics become considerably simpler reducing into a 
system of two coupled-channel integral equations in each case. It is found that, in these reduced systems, the type-A 
elastic channel does not exhibit a limit cycle behavior any longer, while the type-B elastic channels continue to 
exhibit limit cycles but instead following a very different value of the asymptotic parameter, namely, 
$s^{\prime\infty}_0\approx 0.84$, for each mirror system. Subsequently, it may indeed be checked that the estimation 
of the scattering lengths $a_{3(s,t)}$ leads to the expected natures of the Phillips-lines with negative slopes. This
ostensibly indicates plausible role of the simultaneous particle-dimer thresholds resulting in the atypical nature of
the type-B Phillips-lines. However, a more satisfactory explanation of this feature demands a thorough understanding 
of the off-shell dynamics perhaps hinting at the need of a four-body calculations which is beyond the present scope. 
Finally, the fact that our results converge asymptotically for momentum scales significantly larger than $m_\pi$, 
the canonical hard scale of ${}^{\pi\!\!\!/}$EFT, there is an indication of the apparent insensitivity of the three-body
dynamics to the $\Lambda$-$\Lambda$ (two-body) correlations. In this regard, our findings corroborate the two previous 
${}^{\pi\!\!\!/}$EFT analyses~\cite{Ando:2013kba,Ando:2015uda} based on similar three-body calculations of 
${}_{\Lambda\Lambda}^{\,\,\,\,4}{\rm H}$ and ${}_{\Lambda\Lambda}^{\,\,\,\,6}{\rm He}$ double-$\Lambda$-hypernuclei.  
%-----------------------------------------
%-----------------------------------------

%%%%%%%%%%%%%%%%%%%%%%%%%%%%%%%%%%%%%%%%%%%%%%%%%%%%%%%%%%%%%%%%%%%%%%%
\section{SUMMARY AND CONCLUSIONS}\label{sec:summary}
%%%%%%%%%%%%%%%%%%%%%%%%%%%%%%%%%%%%%%%%%%%%%%%%%%%%%%%%%%%%%%%%%%%%%%%

In summary, this work presents an assay of the putative doubly strange ($S=-2$) mirror double-$\Lambda$-hypernuclei
(${}_{\Lambda\Lambda}^{\,\,\,\,5}{\rm H}\,,\,{}_{\Lambda\Lambda}^{\,\,\,\,5}{\rm He}$) in the context of a LO pionless EFT.
In this framework the systems are conjectured as shallow three-particle halo-bound clusters, viz. the iso-doublet pair 
($\Lambda\Lambda t,\,\Lambda\Lambda h$) in the $J=1/2$ channel. The numerical methodology presented here closely resembles 
the approaches of Refs.~\cite{Ando:2013kba,Ando:2015uda,Ando:2015fsa,Raha:2017ahu}. By solving the Faddeev-like coupled 
integral equations~\cite{STM1,STM2,DL61,DL63} for each choice (type-A, -B) of the constituent $(\Lambda T)_{s,t}$ subsystem 
spin introduced in the elastic channel, we presented a qualitative RG based study of the cutoff dependence of the 
three-body contact interactions. In particular, we investigated the dynamical interplay between the different constituent 
two-body subsystems, namely, the virtual bound ${}^1$S${}_0$ $\Lambda\Lambda$ cluster (with $a_{\Lambda\Lambda}<0$), and 
the $(\Lambda T)_{s,t}$ bound clusters (equivalently, the two-body spin-singlet and spin-triplet bound states, i.e., 
${}_{\Lambda}^{4}{\rm H}[J=0^+,1^+]$ and ${}_{\Lambda}^{4}{\rm He}[J=0^+,1^+]$), whose interplay could plausibly lead to the
emergence of three-body shallow bound states. This is formally suggested by the appearance of RG limit cycles in the running
of the three-body couplings $g^{(A,B)}_3(\Lambda_c)$. In the unitary limit this also implies that a discrete sequence of 
Efimov states emerges from the three-particle threshold~\cite{Efimov:1970zz}, and simultaneously with our LO theory in the 
{\it scaling limit}, the ground-state energy collapses to negative infinity (Thomas effect~\cite{Thomas:1935}). Of course, 
such universal phenomena are {\it de facto} unrealistic and disappear for interactions with nonvanishing range (finite 
momentum cutoff) and finite scattering lengths. Nevertheless, for energies in proximity to the particle-dimer thresholds 
(sufficiently far from open channels involving transmutations into particles like $\Sigma,\, \Xi,\cdots$) with reasonably 
fine-tuned $\Lambda$-$T$ and $\Lambda$-$\Lambda$ correlation strengths, it can not be precluded that any remnant universal feature leads to the formation of Efimov-like trimers.

\vspace{0.1cm}

For our numerical analysis we considered different choices of the input double-$\Lambda$ scattering lengths within the 
currently acceptable range, 
$-1.92\,\, {\rm fm} \lesssim a_{\Lambda\Lambda} \lesssim -0.5 \,\, {\rm fm}$~\cite{Morita:2014kza,Ohnishi:2015cnu,Ohnishi:2016elb}, 
along with the inputs for the $\Lambda+u_{0,1}$ particle-dimer thresholds provided by the up-to-date experimental 
information on the $\Lambda$-separation energies of the 
($\!\!\!{}_{\,\,\,\,\Lambda}^{\,\,\,\,4}{\rm H}\,,\!\!\!{}_{\,\,\,\,\Lambda}^{\,\,\,\,4}{\rm He}$) mirror hypernuclei~\cite{Davis:2005mb,Yamamoto:2015avw,Esser:2015trs,Schulz:2016kdc,Koike:2019rrs}. By appropriate fixing of the 
three-body contact interactions using the RG limit cycles at the typical cutoff scale, $\Lambda_c\sim 200$~MeV, a fairly good 
agreement of our EFT predicted $B_{\Lambda\Lambda}\,$-$\,a_{\Lambda\Lambda}$ correlations was obtained with existing potential
models~\cite{Filikhin:2002wm,Filikhin:2003js,Myint:2002dp,Lanskoy:2003ia,Nemura:2004xb} (provided that the 
$B_{\Lambda\Lambda}$ are reevaluated from their old model prediction of $B_{\Lambda\Lambda}$ using updated experimental 
inputs). This agreement, of course, relied on the efficacy in choosing our normalization points, taken from the 
{\it ab initio} potential model analysis of Nemura {\em et al.}~\cite{Nemura:2004xb}. In this case the 
double-$\Lambda$-separation energy $B_{\Lambda\Lambda}$ could be identified with the eigenenergy of the ground ($n=0$) state 
Efimov-like trimer, with the provision that our halo/cluster ${}^{\pi\!\!\!/}$EFT analysis could be extended to include 
$\pi\pi$ or $\sigma$-meson exchange interactions with an adjusted breakdown scale, $\Lambda_H\gtrsim 2m_\pi$. But whether
such physically realizable bound states can be {\it de facto} supported in our EFT framework remains contentious, depending
crucially on support from experimental or lattice QCD data which are currently altogether missing. Future feasibility studies
from the much awaited production experiments, like PANDA and CBM at FAIR~\cite{Pochodzalla:2010,Boca:2015,Vassiliev:2017},
and JPARC-P75~\cite{Fujioka:2019}, are likely to explicate more on the inherent character of these hypernuclei. Besides, 
predictions based on LO EFT analyses are by and large qualitative in nature and must be supplemented by subleading order 
precision analyses for robust assessments. This should naturally address issues such as the compatibility of the low-energy 
cluster picture at momentum scales, $Q \gtrsim \Lambda_H$, potentially probing the short-distance degrees of freedom beyond 
the breakup scales of the triton and helion cores.      

\vspace{0.1cm}

Finally, to demonstrate the predictive power of our EFT analysis, we presented preliminary estimates of the 
$\Lambda$-separation energies ${\mathcal B}_\Lambda$ of the two double-$\Lambda$-hypernuclear mirrors of interest and the 
previously undetermined $S$-wave three-body scattering lengths for the ${}_{\Lambda}^{4}{\rm H}\,$-$\,\Lambda$ and 
${}_{\Lambda}^{4}{\rm He}\,$-$\,\Lambda$ scattering processes. Needless to say that, with the scarcity of pertinent empirical inputs,
a theoretical error analysis based on such empirical estimates serves little purpose and, hence, was not attempted in this work. 
Nevertheless, the accuracy of our results evidently relies on the precise nature of the 
$B_{\Lambda\Lambda}\,$-$\,a_{\Lambda\Lambda}$ correlations, with the latter being still poorly constrained currently. Subject 
to the inherent limitation pertaining to the ambiguity in the normalization of the solutions to the integrals equations, our 
EFT methodology demands a three-body empirical input which is provided by the $B_{\Lambda\Lambda}$ model predictions of Nemura 
{\it et al}. Subsequently, the correlation plots self-consistently determine the three-body scattering lengths 
$a_{\Lambda\Lambda T}$. In particular, the scale variation of the renormalized scattering lengths was found to asymptotically 
converge for $\Lambda_c\gtrsim 500$~MeV which is well beyond the hard scale of standard ${}^{\pi\!\!\!/}$EFT. Thus, the 
three-body dynamics are most likely insensitive to the low-energy $\Lambda$-$\Lambda$ two-body interactions, unless the hard-scale $\Lambda_H$ of the effective theory could be augmented sufficiently beyond without potentially invalidating the basic 
halo/cluster {\it ansatz}. This supports the earlier claim made in Refs.~\cite{Ando:2013kba,Ando:2015uda} based on similar 
investigations of the other double-$\Lambda$-hypernuclear cluster systems, such as ${}_{\Lambda\Lambda}^{\,\,\,\,4}{\rm H}$ 
and ${}_{\Lambda\Lambda}^{\,\,\,\,6}{\rm He}$. Although short-distance mechanisms beyond the realm of our EFT can certainly 
influence the formation of such exotic bound hypernuclear clusters, this does not preclude possible role of low-energy 
off-shell effects that may not be accessible in a three-body framework without involving four-body calculations. Such an 
endeavor, however, goes beyond the scope of the simple qualitative nature of this work. To this end, we reiterate once more 
that our estimates of the scattering lengths $a_{\Lambda\Lambda T}$ should serve for demonstrative purposes only, given the 
current limitations of performing $\Lambda\Lambda T$ scattering experiments in testing their validity thereof. \\

%%%%%%%%%%%%%%%%%%%%%%%%%%%%%%%%%%%%%ACKNOWLEDGEMENT%%%%%%%%%%%%%%%%%%%%%%%%%%%%%%%%%%%%%%%%%%%%%%%%
\section*{Acknowledgments}
%%%%%%%%%%%%%%%%%%%%%%%%%%%%%%%%%%%%%%%%%%%%%%%%%%%%%%%%%%%%%%%%%%%%%%%%%%%%%%%%%%%%%%%%%%%%%%%%%%%
We are thankful to S.-I. Ando for providing many useful suggestions regarding this work. We are also 
thankful to A. Gal for apprising us of the pioneering work of Ref.~\cite{Contessi:2019csf} on the 
{\it ab initio} (SVM) pionless EFT analysis regarding the onset of double-$\Lambda$-hypernuclear 
binding.

%%%%%%%%%%%%%%%%%%%%%%%%%%%%%%%%%%%%%%%%%%%APPENDIX%%%%%%%%%%%%%%%%%%%%%%%%%%%%%%%%%%%%%%%%%%%%%%%%%
\section{Appendix}
%%%%%%%%%%%%%%%%%%%%%%%%%%%%%%%%%%%%%%%%%%%%%%%%%%%%%%%%%%%%%%%%%%%%%%%%%%%%%%%%%%%%%%%%%%%%%%%%%%%%

%%%%%%%%%%%%%%%%%%%%%%%%%%%%%%%%%%%%%%%%%%%%%%%%%%%%%%%%%%%%%%%%
\subsection{One- and Two-body non-relativistic Propagators}
\begin{figure}[bp]
    \centering
    \includegraphics[width=\columnwidth]{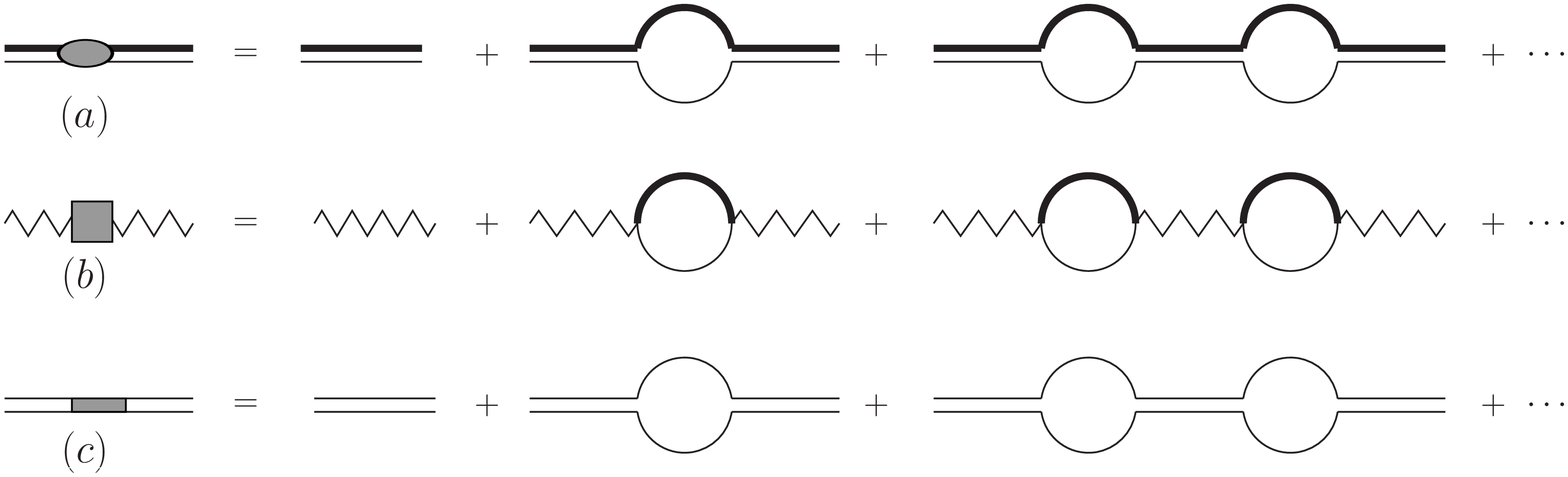}
    \caption{\label{fig:fig10} Diagrams for the renormalized dressed dimer propagators: 
             (a) $i\Delta_0$ for the spin-singlet auxiliary field $u_0$, 
             (b) $i\Delta_1$ for the spin-triplet auxiliary field $u_1$, and 
             (c) $i\Delta_s$ for the spin-singlet auxiliary field $u_s$. 
             Thick (thin) lines denote the $\Lambda$-hyperon (core $T\equiv t,h$) field propagators. }
\end{figure}

%%%%%%%%%%%%%%%%%%%%%%%%%%%%%%%%%%%%%%%%%%%%%%%%%%%%%%%%%%%%%%%%
Here we summarize the one- and two-body nonrelativistic propagators specific to the 
$\Lambda\Lambda T$ three-body systems in pionless effective theory (${}^{\pi\!\!\!\!/}$EFT). 
In this framework, at sufficiently low-energies below the respective breakup scales, we
may consider the triton (${}^3$H or $t$) and the helion (${}^3$He or $h$) as being 
fundamental particles. Thus, as the fundamental one-body components of the theory, the 
$\Lambda$ and $T$ propagators are given as
\begin{eqnarray}
 iS_{\Lambda,\,T}(p_0,{\bf p})&=&\frac{i}{p_0-\frac{{\bf p}^2}{2M_{\Lambda,\, T}}+i\eta}\,,
\end{eqnarray}
where $p_0$ and ${\bf p}$ are the generic off-shell energy and three-momentum. In our analysis we only 
consider the $S$-waves contributions from the two-body interactions at LO. We have incorporated a power 
counting scheme~\cite{Kaplan:1998tg,Kaplan:1998we} for the 
${}^{1}{\rm S}_0$ $\Lambda$-$T$, ${}^{3}{\rm S}_1$ $\Lambda$-$T$ and the ${}^{1}{\rm S}_0$ 
$\Lambda$-$\Lambda$ interactions in the two-body sector, in which the unitarized two-body amplitudes are 
conveniently expressed in terms of the auxiliary fields, namely, the spin-singlet and spin-triplet 
$\Lambda T$-dimer fields $u_{0,1}$, and the spin-singlet $\Lambda\Lambda$-dibaryon field $u_s$. The leading 
order renormalized dressed dimer propagators~\cite{Bedaque:1998kg,Bedaque:1998km,Bedaque:1999ve} are 
given by the expressions (see Fig.~\ref{fig:fig10}\,)
%-----------------------------------------
%-----------------------------------------
\begin{eqnarray}
 {\mathcal D}_{0,1}(p_0,{\bf p})&=&
\frac{1}{\gamma_{0,1}-\sqrt{-2\mu_{\Lambda T}(p_0-\frac{ {\bf p}^2}{2(M_T+M_\Lambda)})-i\eta}-i\eta}\,\,,
\nonumber\\
\end{eqnarray}
and,
\begin{eqnarray}
{\mathcal D}_{s}(p_0,{\bf p})&=&
\frac{1}{\frac{1}{a_{\Lambda\Lambda}}-\sqrt{-M_\Lambda(p_0-\frac{ {\bf p}^2}{4M_\Lambda})-i\eta}-i\eta}\,,\quad\,
\label{dimer_props}
\end{eqnarray}
respectively, with the LO two-body contact interactions $y_{0}$, $y_{1}$, and $y_{s}$ fixed as in Eq.~\eqref{eq:y01s} in 
the text. In the above expressions, $\gamma_{0}$ and $\gamma_{1}$ are the binding momenta of spin-singlet and spin-triplet 
states of the $\Lambda T$ subsystem, and $a_{\Lambda\Lambda}$ is $S$-wave double-$\Lambda$ scattering length. 


%
%\bibliographystyle{apsrev4-1}
%\bibliographystyle{h-physrev4}
%\bibliography{refs}

%\newpage

%%%%%%%%%%%%%%%%%%%%%%%%%%%%%%%%%%%%%%%%%%%%%%%%%%%%%%%%%%%%%%%%%%%%%%%

\end{document}